# Electronic Transport in Two-Dimensional Materials

*Vinod K. Sangwan[1] and Mark C. Hersam[1,2,3,*]*

[1]Department of Materials Science and Engineering, Northwestern University, Evanston, Illinois 60208, USA.

[2]Department of Chemistry, Northwestern University, Evanston, Illinois 60208, USA.

[3]Department of Electrical Engineering and Computer Science, Northwestern University, Evanston, Illinois 60208, USA.

*Corresponding author e-mail: m-hersam@northwestern.edu

**Abstract**

Two-dimensional (2D) materials have captured the attention of the scientific community due to the wide range of unique properties at nanometer-scale thicknesses. While significant exploratory research in 2D materials has been achieved, the understanding of 2D electronic transport and carrier dynamics remains in a nascent stage. Furthermore, since prior review articles have provided general overviews of 2D materials or specifically focused on charge transport in graphene, here we instead highlight charge transport mechanisms in post-graphene 2D materials with particular emphasis on transition metal dichalcogenides and black phosphorus. For these systems, we delineate the intricacies of electronic transport including bandstructure control with thickness and external fields, valley polarization, scattering mechanisms, electrical contacts, and doping. In addition, electronic interactions between 2D materials are considered in the form of van der Waals heterojunctions and composite films. This review concludes with a perspective on the most promising future directions in this fast-evolving field.

## 1. Introduction

Over the past decade, the superlative properties of single-layer graphene have inspired aggressive exploration of other 2D materials, initially focusing on transition metal dichalcogenides,[1-10] but more recently broadening to the larger family of layered materials.[11-18] Among the unique physical and chemical characteristics of 2D materials, electrical and optical properties present specific opportunities in electronic and optoelectronic devices including transistors, photodetectors, light-emitting diodes, and solar cells. Indeed, the most optimistic projections view atomically thin 2D materials as a post-silicon alternative as conventional electronics approach the limits of traditional scaling.[5,11,19-21] Although electronic transport in bulk layered semiconductors have been studied for over four decades,[22-24] the effects of reduced screening and quantum confinement in the 2D limit are only beginning to be unraveled.[25-27] For example, theoretical and experimental efforts have identified intrinsic carrier mobility limits for 2D materials on idealized atomically flat substrates,[17,28,29] but the extension of these results to realistic, wafer-scale geometries remains in the early stages of development.[30]

The weak van der Waals (vdW) interactions that enable facile exfoliation of atomically thin sheets from layered solids also allow the direct assembly of heterojunctions consisting of distinct 2D materials.[31-36] 2D materials can also be integrated with other electronic materials including zero-dimensional (0D) quantum dots, one-dimensional (1D) nanotubes, and bulk semiconductors to achieve mixed-dimensional vdW heterostructures with unprecedented device behavior.[5,35,37-39] While these recently emerging heterojunctions are reminiscent of more traditional compound semiconductor heterostructures,[40] the underling vdW interactions relax conventional lattice matching constraints, enabling relatively strain-free and charge-free heterointerfaces for most 2D materials.[33,36,38] In this manner, a wide array of vdW heterostructures can be realized with broad potential utility in applications such as photodetectors, photovoltaic cells, and aggressively scaled digital circuits.[19,20,35,41,42] However, compared to highly developed compound semiconductors, 2D materials typically show inferior electronic properties including lower mobility, reduced stability, and higher contact resistances that require further study and optimization before they can effectively compete in real-world technology.[30,43-45]

Here, we provide an overview of the factors that underlie electronic transport and ultimately control device performance in 2D materials including bandstructure, screening, scattering, doping, and contacts. The relative importance of these issues is assessed through state-of-the-art metrics and implications for new device architectures. Since pre-existing review articles on 2D materials have been dedicated to materials surveys,[6,8] devices applications,[5,10,20,46] growth and processing,[7,47,48] chemistry,[49] and van der Waals heterojunctions,[33,36,38,50] these topics will not be the principal focus here. We will also not discuss electron-electron interactions, metal-insulator transitions, or correlated electron phenomena such as charge-density waves (CDW) and superconductivity since they have been reviewed previously.[51,52] Instead, this review article is principally dedicated to electronic transport in 2D materials, and aims to serve as an interdisciplinary tutorial and roadmap for future work in this field.

## 2. Introduction to 2D materials

The experimental isolation of monolayer graphene initiated voluminous activity in condensed matter physics to study the exotic properties resulting from its linear dispersion and massless Dirac Fermions.[53,54] In addition, the large room-temperature electron mobility in graphene (>50,000 $cm^2$/Vs) naturally led to applications in high-speed electronic devices.[4,28]

However, the zero bandgap of graphene presents significant issues in most electronic applications, which inspired subsequent efforts to modify graphene in an effort to open a bandgap. For example, a bandgap can be opened in graphene by breaking the symmetry of K and K' points in the first Brillouin zone, which led to explorations of graphene-related materials, such as graphene oxide and hexagonal boron nitride (hBN). In this context, hBN is an extreme example where replacing carbon in the K and K' positions with boron and nitrogen results in a wide bandgap (~6 eV) insulator. By providing a nearly defect-free atomically flat dielectric interface, hBN can be effectively mated with other 2D materials to reveal nearly intrinsic properties,[4,53,54] which will be discussed in more detail below.

Beyond the graphene family, the most studied 2D materials are the layered transition metal dichalcogenide (TMDC) compounds, $MX_2$, where M is a transition metal (e.g., Mo, W, Re, and Ta) and X is a chalcogen (e.g., S, Se, and Te). Following the isolation of monolayer graphene, more than 30 stable TMDCs have been studied using similar methods over the past decade.[6,8,24] Preceding the recent interest in the 2D limit, bulk layered TMDCs were extensively studied for numerous applications in catalysis and lubrication. It should be noted that this early work included the first isolation of monolayer $MoS_2$ using adhesive tape assisted mechanical exfoliation in 1966 and by chemical exfoliation 20 years later.[55,56] Following exfoliation, the monolayer thickness of ~0.7 nm was verified by measuring the width of the shadow created by metal evaporated at an angle.[55] Later work focused on Hall measurements, which resulted in theories for electron-phonon scattering in TMDCs such as $MoS_2$ and $WS_2$.[22,23] Around the same time, Group V TMDCs such as $TaS_2$ and $NbSe_2$ were also explored for strongly correlated electron phenomena such as charge density waves and superconductivity.[52]

Typical crystal structures of monolayer TMDCs include trigonal prismatic 1H (2H for multilayer), 1T, and 1T' (i.e., distorted-1T) phases (Fig. 1a–c).[8,24] The electronic structure of TMDCs depends sensitively on the crystal phase, resulting in a range of electronic character including metallic, semimetallic, semiconducting, and superconducting for different TMDCs.[52,57] TMDCs based on Groups V and VI tend to be the most heavily studied due to the diverse permutations of stable compounds and electronic behavior.[5,6,20] Indeed, stable compounds with all permutations between Group V and VI metals (i.e., V, Nb, Ta, Cr, Mo, and W) and chalcogens (i.e., S, Se, and Te) exist except $CrS_2/Se_2/Te_2$ and $VS_2$.[8,24] Layered sulfides of Group IV (i.e., Ti, Hf, and Zr) also possess interesting semimetallic and semiconducting behavior, but they suffer from a high propensity for non-stoichiometric structure due to low energy barriers for intercalation of metal atoms.[24,58] Group V and VII TMDCs are mostly studied for many-body phenomena such as 1T $TaS_2$ for gate-tunable charge density waves and 1T' $ReS_2$ for linear anisotropy in electrical and optical properties.[2,6,8,18,24,59] Chalcogenides and halides for Group VIIIa,b,c transition metals (e.g., pyrites and marcasites) show both layered and non-layered structures with equally diverse electronic structures (e.g., $CdI_2$-structure family) including semiconductors such as $FeS_2$.[6,8,24]

In the 2D limit, semiconducting TMDCs, specifically $MoS_2$, $WS_2$, and $WSe_2$, have emerged as attractive materials for electronic devices due to their relatively high charge carrier mobilities and appreciable bandgaps that enable large switching ratios in field-effect transistors (FETs).[5,6,20,60] Furthermore, successful growth at the wafer-scale and ambient stability increase their prospects for practical applications.[30] With direct bandgaps at the monolayer limit, large oscillator strengths, large spin-orbit splitting, and access to the valley degree of freedom at room temperature, TMDCs also represent an interesting platform for fundamental studies of light-matter interactions, optoelectronics, and nanophotonics.[3,46,61-63] In addition, since bulk TMDCs were historically studied in the catalysis community for desulfurization reactions, the recently renewed interest in 2D TMDCs has reinvigorated the study of this class of materials in the fields of catalysis, chemical and biological sensing, and energy storage.[9,49,60]

Layered post-transition metal chalcogenides (PTMCs), such as GaS, SnS, $SnS_2$, and InSe (Fig. 1d), are also being explored due to their high mobilities, large photoresponsivities, and in-plane anisotropy.[17,64,65] While some layered PTMCs (e.g., $Bi_2Se_3$, see Fig. 1e) have been long studied as thermoelectric materials, more recent work has focused on their potential as topological insulators (i.e., topologically-protected surface states are conducting while the bulk remains insulating).[66] Layered dihalides and trihalides of transition metals (e.g., $FeCl_2$ and $CrI_3$) and post-transition metals (e.g., $PbI_2$ and $BiI_3$) also show exotic optical and magnetic behavior such as ferromagnetism in an insulating state.[67,68] However, most of these materials are not compatible with semiconductor processing, which implies that their applications in solid-state devices are limited. Layered 2D oxides such as $MoO_3$ (Fig. 1f),[69,70] 2D carbides or MXenes such as $Mo_2C$,[71] and perovskites[72] are also being explored for energy storage, catalysis, and photovoltaics. Layered transition metal hydroxides (e.g., $Mn(OH)_2$) and double hydroxides, mostly synthesized *via* wet chemistry routes, are finding niche applications in energy storage, $CO_2$ splitting, and hydrogen evolution.[70] Layered materials based on actinides ($USe_2$) and rare earth metals ($GdCl_3$) have also been isolated and characterized, but their use is not widespread.[24]

In addition to 2D compounds, significant effort has been devote to 2D elemental materials such as B (borophene), Si (silicene), P (phosphorene), As (arsenene), Sb (antimonene), Bi (bismuthene), and Sn (stanene).[14,73] However, most of these elemental monolayers (e.g. borophene, see Fig. 1g) are grown on metal surfaces in ultra-high vacuum (UHV) and show high chemical reactivity in ambient conditions, thus limiting device applications.[16] Thus far, silicene[15] and black phosphorus (BP)[11,12] have proven to be the most promising electronic materials in this class. BP is most actively being pursued due to its high field-effect mobility (~1000 cm$^2$/Vs at room temperature), direct and tunable bandgap at all thicknesses, and large in-plane anisotropy due to its puckered atomic structure (Fig. 1h).[11,12,49] Furthermore, easily achieved p-type conduction in BP complements n-type TMDC semiconductors in digital logic and p-n heterojunction applications. Since TMDCs and BP have emerged as the leading 2D materials for electronic applications, they have received the most attention in research devoted to charge transport. Therefore, we will focus most of our subsequent discussion on these 2D materials, particularly using MoS$_2$ and BP as model systems.

## 3. Tunable bandstructure with thickness and fields

For comparative purposes, Fig. 2 shows the band edges and bandgap for selected monolayers of 2D materials along with conventional semiconductors and metals. Due to strong quantum confinement in the vertical direction, the electronic structure of 2D materials can change significantly with the number of layers.[3,17,25] Size-dependent bandgaps and plasmon frequencies have been quintessential features of most low-dimensional materials such as quantum dots (QDs) and carbon nanotubes (CNTs).[74] Group VI TMDCs (e.g., MoS$_2$, MoSe$_2$, WS$_2$, and WSe$_2$) in 2H phases show an evolution from an indirect bandgap of ~1.3, 1.1, 1.4, 1.2 eV in the bulk to a direct bandgap of ~1.9, 1.6, 2.1, 1.7 eV in the monolayer limit, respectively.[6,27,75] As the thickness is increased from monolayer to bilayer, the direct bandgap at the K point of the Brillouin zone remains relatively unchanged due to the dominance of transition metal $d$-orbital that is relatively isolated from interlayer coupling.[25,76] However, the bands at the Γ point change significantly due to the increased contribution of chalcogen $p_z$ orbitals (Fig. 3a). This emergence of a smaller indirect gap between the Γ and Q points in bilayers decreases photoluminescence (PL) quantum yields in MoS$_2$ by up to a factor of ~$10^4$ between monolayers and bilayers.[3] BP also shows a thickness-dependent bandgap from 0.3 eV in the bulk to 1.9 eV in the monolayer limit, although the bandgap remains direct for all thicknesses of BP.[12,13] Recently, the 2D semiconductor InSe (mobility ~1000 cm$^2$/Vs at room temperature) has also shown unusual layer-dependent behavior with direct bandgaps ranging from 1.3 eV in the bulk to 1.8 eV for trilayers.[17] However, below three layers, InSe transitions to an indirect bandgap semiconductor. In particular, in monolayer InSe, the fundamental transition across the bandgap is forbidden due to a combination of out-of-plane mirror symmetry and large $p_z$ character at the band edges. Interestingly, monolayer InSe still shows a PL peak at ~2.9 eV, resulting from recombination between deeper bands (Fig. 3b).[17] Another interesting case is ReS$_2$, which shows weak interlayer coupling, resulting in minimal changes in bandgap from 1.42 eV in the bulk to 1.52 eV in the monolayer limit. Furthermore, the weak interlayer coupling implies that bulk ReS$_2$ effectively behaves as if it is composed of isolated monolayers.[18]

The ultrathin body of 2D materials implies incomplete screening of externally applied electric and magnetic fields, enabling field-dependent tuning of electronic structure and thus novel field-effect devices.[76,77] For instance, PL measurements under vertical gate fields have shown quantum-confined Stark effect in monolayer MoS$_2$ due to the bandgap tuning in bilayer MoS$_2$ and the emergence of interlayer excitons.[77,78] *In situ* scanning tunneling spectroscopy of gated devices also show that the Fermi level in monolayer MoS$_2$ can be moved by up to 0.6 eV in a field-effect transistor geometry.[79] However, the bandstructure near different valleys varies differently with field.[76] For example, atomic-level calculations based on Wannier functions in few-layer TMDCs predict larger field dependence of K-valleys than Γ-valleys or Q-valleys, which are consistent with gated PL measurements (Fig. 3c, d).[27] It should be noted that these experimental PL peak shifts can only be reconciled with the presence of intralayer excitons (i.e., electron-hole pairs within the same layer) as opposed to interlayer excitons (i.e., electron-hole pairs in different layers), which is also consistent with other experiments on layer-independent exciton polarizability.[27,77] External electric field have also been shown to induce charge density wave phase transitions in metallic NbSe$_2$ and TaS$_2$.[52,59]

Electronic structure can also be tuned by reversible switching between metastable phases in TMDCs. In particular, the semiconducting 1H/2H phase of TMDCs (Fig. 1a) can be converted to the metallic 1T phase (Fig. 1b) by Li ion intercalation, after which the 2H phase can be recovered by thermal annealing.[57] Overall, the diversity of electronic properties in 2D materials, including metals and insulators,[52,70] combined with the tunability of semiconducting bandstructure with thickness and fields offer numerous opportunities for electronic devices.[5,6,20,28] However, the strong thickness dependence also presents a challenge in that atomic-level precision is needed in film uniformity for reliable technology at the wafer-scale.[30]

## 4. Valley polarization and anisotropy

In trigonal prisms of TMDCs, a hexagonal plane of transition metal is sandwiched between two hexagonal planes of chalcogens (Fig. 1a, top). Looking from the top surface, alternating K and K' points are occupied by one transition metal and two chalcogens alternatively (Fig. 1a, bottom). This lack of inversion symmetry lifts the K-K' degeneracy, and combined with time reversal symmetry, couples the valley degree of freedom with spin.[62] Furthermore, the valence band incurs large spin-orbit coupling (~0.3 eV),[3] resulting in a scenario where different valleys can be populated by circularly polarized light. Different valleys can also be populated by transverse electric fields, which leads to the valley Hall effect without the need for an external magnetic field.[62,80] It should be noted that the K-K' symmetry is regained in even layered samples.

Early research in layered bulk materials showed large anisotropy between in-plane mobility ($\mu_x$, $\mu_y$) and out-of-plane mobility ($\mu_z$) (e.g., $\mu_x/\mu_z \sim 10^3$ in bulk $MoS_2$). This anisotropy is not necessarily a direct consequence of anisotropy in electronic structure (i.e., effective mass), but rather comes from interlayer van der Waals gaps acting as tunneling barriers.[81] On other hand, in-plane anisotropy in mobility is usually a direct consequence of anisotropy in effective mass. For example, BP and $ReS_2$ show $\mu_x/\mu_y$ ratios of 1.8 and 1.56, respectively (Fig. 3e).[82,83] Similarly, strong in-plane anisotropy has been observed in optical properties, photocurrent, photoluminescence, and thermal conductivity, which is potentially useful for polarization-dependent electro-optical and electro-thermal devices.[13]

## 5. Electronic transport

In this section, we first discuss intrinsic mechanisms affecting electrical conductivity in individual 2D materials, followed by performance-limiting extrinsic factors such as contacts, substrates, and doping in practical devices. Separate sub-sections allow focused discussion and easy access to references, but the sub-sections are interrelated and provide a holistic picture when considered in aggregate.

### 5.1 Carrier screening, scattering, and mobility

The general properties of free electron screening in 2D are well understood from one of the earliest examples of a 2D system, namely the field-effect inversion layer in silicon.[74] More recently, anomalous screening effects in graphene due to its linear energy dispersion proved critical in understanding the origins of high mobility in graphene.[53,54] Fig. 4a shows major sources of carrier scattering in 2D FETs: (1) electron-phonon interactions; (2) Coulomb scattering from charged impurities (CI); (3) surface optical (SO) phonons (also called remote interfacial phonons); (4) surface roughness; and (5) structural defects.[74] Historically, Fivaz and Mooser first studied bulk TMDCs in the 1960s and reported the temperature dependence of the Hall mobility as $\mu = \mu_0 (T/300 \text{ K})^{-\gamma}$ with $\gamma = 2\text{–}3$ dominated by electron-phonon scattering at room temperature ($\gamma = 2.6$, 2.5, 2.4, 2.1 for $MoS_2$, $MoSe_2$, $WS_2$, and GaSe, respectively).[23,84] Similarly, Keyes observed $\mu \sim T^{-1.5}$ trend for mobility in bulk BP in 1953 and justified the same by electron-phonon scattering.[96] More recently, Kassbjerg $et\ al.$ studied electron-phonon scattering in monolayer $MoS_2$ from first principles and calculated a phonon-limited room temperature mobility of ~400 $cm^2$/Vs.[26] Most studies on $MoS_2$ transistors report significantly lower values, which indicates that other scattering mechanism need to be considered.

Models based on Thomas-Fermi (TF) screening capture the most essential features of $MoS_2$ transistors (TF screening length is defined as $\lambda_{TF} = \sqrt{(\varepsilon_{2d} t_{2d} t_{ox})/\varepsilon_{ox}}$ ~7 nm, where $\varepsilon_{2d}$ and $\varepsilon_{ox}$ are dielectric constants of the 2D material and oxide dielectric, and $t_{2d}$ and $t_{ox}$ are thicknesses of the 2D material and oxide dielectric, respectively).[85,86] Competing roles of screening and interlayer resistance were further revealed in uneven current density (hot spots) within few-layer $MoS_2$ (Fig. 4b).[85] More recently, Lindhard screening has been used to fully describe screening of free carriers and CIs, allowing the temperature and carrier density dependence of mobility to be determined.[87] In this manner, the dielectric functions and temperature-dependent polarizabilities have been worked out to obtain scattering matrix elements and momentum relaxation rates for most Group VI TMDCs and BP.[88-91] In particular, the mobility from individual scattering mechanisms ($\mu_i$) is calculated within the relaxation time approximation of the Boltzmann transport equation, after which these terms are added *via* Matthiessen's rule ($\mu_{net}^{-1} = \sum \mu_i^{-1}$).[26,74,76,87] For TMDCs, two classes of electron-phonon interactions appear to be dominant: (1) lattice deformation potential (DP), which involves quasi-elastic scattering by longitudinal acoustic (LA) and transverse acoustic (TA) phonons; (2) Fröhlich interaction for inelastic scattering by in-plane optical phonons and out-of-plane homopolar phonons. The resulting scattering matrix from Fermi's golden rule is divided by the effective dielectric constant to obtain the relaxation rate for different scattering processes.[74,87]

Due to the ultrathin body of 2D materials, the dielectric mismatch with surrounding media can significantly alter the shape of the Coulomb potential within the 2D semiconductor.[87] In Fig. 4c, $q_{TF}/q = \varepsilon_{2d} - 1$, where $\varepsilon_{2d}$, $q_{TF}$, and $q$ are the dielectric constant, TF screening wave vector, and phonon wavevector, respectively, is plotted against effective dielectric constant of the environment, $\varepsilon_e = (\kappa_1 + \kappa_2)/2$ from Fig. 4a.[87] For low-$\kappa$ environments (i.e., $\varepsilon_e < \varepsilon_{2d}$), screening is stronger for low scattering angles, whereas for high-$\kappa$ environments (i.e., $\varepsilon_e > \varepsilon_{2d}$), the opposite is true. Therefore, free carrier screening is weakened by high-$\kappa$ environments, and CI screening is enhanced by high-$\kappa$ environments. High-$\kappa$ materials (e.g., $HfO_2$ and $ZrO_2$) also have low-energy SO phonons that can be excited remotely by electrons in the semiconductor. Although high-$\kappa$ dielectrics have enhanced mobility in graphene, the results are mixed in the case of TMDCs. A systematic study on different combinations of top-gate and bottom-gate dielectrics ($\varepsilon_e = 1 - 25$) concluded that most high-$\kappa$ dielectrics are likely to degrade mobility in $MoS_2$ (Fig. 4d,e).[87] Therefore, the best strategy to maximize mobility is to reduce CI scattering using a suspended geometry or hBN dielectrics. As expected, the mobility of 2D materials increases with the number of layers for the first few layers due to improved screening of CIs and then begins to decrease in thicker samples due to increased interlayer resistance.[86] Extending this concept, mobility can be improved by introducing a spacer such as PMMA between the 2D semiconductor channel and the oxide substrate.[92] The dependence of mobility on carrier density $n_{2d}$ can also be non-trivial. For example, at low $n_{2d}$, mobility increases due to increased screening of CIs by free carriers, but at high $n_{2d}$, the characteristic energy levels of carriers is pushed closer to the polar LO phonon that scatters by the Fröhlich interaction. Thus, the transfer characteristics of $MoS_2$ FETs are typically nonlinear with the drain current $I_d$ varying as $\sim(V_g)^m$, where m = 2 for no screening and m = 1 for complete screening, with the mobility in thicker samples saturating at a smaller $V_g$.[93]

The ultrathin body of 2D materials also makes them highly sensitive to surface roughness ($\Delta L$). Surface roughness is also a common problem in conventional III-V semiconductor heterojunctions where the mobility degrades as the 'sixth-power law' with surface roughness ($\Delta L^{-6}$) due to perturbation of quantized energy levels in the 2D electron gas.[74] This effect is even more significant in high-mobility 2D materials such as graphene, which implies that intrinsic mobility limits are achieved on the atomically flat surface of hBN. Structural disorder such as point defects and grain boundaries can also have greater deleterious effects in 2D materials compared to bulk semiconductors due to larger scattering cross-sections. Specifically, point defects typically cause short-range scattering, whereas charged defects and grain boundaries induce long-range fields for Coulomb scattering. On other hand, sulfur vacancies act as dopants that can improve mobility in CVD $MoS_2$ by filling up traps states below the mobility edge.[43]

The intrinsic limits of mobility in 2D materials have been achieved by careful control of substrate and contact engineering.[94] For example, few-layer $MoS_2$ sandwiched between hBN films and contacted by graphene (Fig. 5a) shows low-temperature mobility exceeding $10^4$ cm$^2$/Vs (limited by CIs) and show Shubnikov–de Haas oscillations, while room-temperature mobility (~100 cm$^2$/Vs) is limited by optical phonons.[29] The temperature exponent ($\mu \sim T^{-\gamma}$) ranges between 1.9 and 2.3, comparable to the optical phonon-limited $\mu \sim T^{-2.6}$ for bulk MoS$_2$ (Fig. 5b). Few-layer (monolayer) MoS$_2$ devices on SiO$_2$ show comparable room temperature mobilities of ~120 cm$^2$/Vs (~60 cm$^2$/Vs) but significantly reduced low temperature mobilities of ~400 cm$^2$/Vs (~120–300 cm$^2$/Vs) with $\gamma = 0.67$–1.7 in unencapsulated devices and $\gamma = 0.3$–0.73 in top-gated devices.[95-99] Since a $\mu \sim T^{-1}$ trend is expected from acoustic deformation potential scattering above the Bloch-Grüneisen temperature,[26] the measured values of $\gamma$ are justified in first principles mobility models for monolayer MoS$_2$ by considering Raman-active optical phonons (E$^1_{2g}$ ~49 meV) and homopolar phonons (A$_{1g}$ ~52 meV) (Figs. 4e).[26,87,100] These mobility models also explain the ~50% decrease in $\gamma$ (i.e., ~37% increase in μ) in top-gated devices.[100] Sandwiching schemes based on hBN have also been used to probe the intrinsic performance of other 2D semiconductors such as hBN-sandwiched InSe yielding low-temperature mobilities of ~10$^4$ cm$^2$/Vs (and Shubnikov–de Haas oscillations) and room-temperature mobilities of ~1000 cm$^2$/Vs, which are the highest for any n-type 2D semiconductor. Among p-type materials, few-layer BP has shown the highest mobility values (~1000 cm$^2$/Vs at room temperature), although the bandgap of few-layered BP is essentially that of bulk BP (~0.3 eV), which leads to a modest switching ratio (< 500). Finally, variable range hopping has been reported in some low-mobility MoS$_2$ transistors ($\mu$ < 10 cm$^2$/Vs).[93,98] The conflicting reports on hopping *versus* diffusive transport for MoS$_2$ suggests that processing residues and imperfect electrical contacts play an important role in many cases.[93,96,98]

It is instructive to compare the mobility and bandgap of the most promising 2D materials with competing technologies such as thin-film transistors (TFTs), silicon-on-insulator (SOI), and ultra-thin body (UTB) devices (Fig. 5c).[5,20] Due to its high-mobility, large switching ratio (>$10^8$), and reduced short-channel effects, the progress in $MoS_2$ transistors has been monitored by ITRS since 2012.[21] In particular, mechanically exfoliated $MoS_2$ devices show comparable mobility to commercial UTB devices on strained-Si,[101] Si-Ge alloys,[102] and InGaAs-on-insulator.[103] However, the performance of large-area $MoS_2$ devices is thus far inferior, which implies that they are better suited as an alternative to TFT technology that is currently dominated by polycrystalline Si, organic semiconductors, and metal-oxides such as InGaZnO (Fig. 5c).[30,43,104-106] Multilayer BP and InSe flakes show higher mobility than UTB technology, but the issue of large-area growth and environmental stability has to be resolved for practical viability. Furthermore, although multilayer devices show higher mobility and are more forgiving for non-uniform large-area growth, they compromise short-channel effects that are considered to be among the most important advantages of 2D electronic devices.

## 5.2 Short-channel effects and high frequency devices

Since all dimensions ideally scale proportionally in short-channel devices, 2D materials offer the ultimate channel length ($L_{ch}$) limit for integrated circuits.[21] At short $L_{ch}$, field-effect mobility is no longer the most relevant metric of performance because transport is ballistic. Instead, current increases with effective mass $m^*$ by influencing the density of states $g_{2d}$ =$g_v m^*/\pi\hbar^2$, where $g_v$ is valley degeneracy and $\hbar$ is Plank constant.[74] Furthermore, at sub-5 nm channel lengths, few-layer $MoS_2$ is expected to have enhanced electrostatic control from the gate due to a lower in-plane dielectric constant (~4) than Si (~11.7) or GaAs (~12.9).[40,107] Most Group VI TMDCs are also expected to have smaller drain leakage current than Si and GaAs due to larger $m^*$ (for $MoS_2$, $MoSe_2$, $MoTe_2$, $m^*$ ~0.56–0.66 $m_0$) compared to Si ($m^*$ ~0.29 $m_0$) and GaAs ($m^*$

~0.15 $m_0$).[25,27,40,76] It should be noted, however, that the saturation velocity ($v_s$) in MoS$_2$ (2.8 × 10$^6$ cm/s) is somewhat smaller than Si (~10$^7$ cm/s), GaAs (7.2 × 10$^6$ cm/s), and graphene (5.5 × 10$^7$ cm/s), but still sufficiently high to compete with alternative UTB technologies.[40,108]

The critical value of $L_{ch}$ ($L_c$) where the gate begins to lose complete control over the channel is defined as $L_c = \alpha\sqrt{(\varepsilon_{2d}t_{2d}t_{ox})/\varepsilon_{ox}}$, where $\alpha \approx 4-6$ for FETs.[109] In one study, an MoS$_2$ FET ($t_{2d}$ = 5 nm, $t_{SiO2}$ = 300 nm) showed complete switching (on/off ratio ~10$^9$) at $L_{ch}$ = 100 nm,[110] while another study on varying $L_{ch}$ and $t_{2d}$ ($t_{SiO2}$ = 90 nm) showed the expected parabolic relation between $L_c$ and $t_{2d}$, such that a 5 nm thick MoS$_2$ FET could not be turned off even at $L_{ch}$ = 200 nm.[109] Recently, it was shown that MoS$_2$ FETs could be turned off by an embedded gate defined by a single carbon nanotube (diameter, $L_g$ ~1 nm), although the total channel length was actually ~500 nm so the device was not formally in the short-channel regime.[107]

In high frequency applications, MoS$_2$ FETs with gate length $L_g \sim L_{ch}$ = 500 nm showed moderate current saturation and a cut-off frequency ($f_T = v_s/(2\pi L_g)$ of 6.7 GHz.[111] Meanwhile, higher mobility BP has shown superior high frequency performance in scaling and speed. For example, 20 nm long BP FETs showed on/off ratios of ~100 with width-normalized currents exceeding 0.1 mA/μm.[42] Similarly, radio frequency BP devices have been realized with $f_T$ ~ 20 GHz at $L_g$ = 300 nm.[112] Further improvements in amplifier speed would likely be enabled by higher mobilities and shorter $L_g$, preferably using the ultraclean interfaces of hBN sandwiches. For high output impedance in practical circuits, the current saturation also needs to be improved in short-channel devices.

### 5.3 Electrical contacts

Electrical contacts play the integral role of charge injection and collection in all electronic devices. The quality of contacts in transistors is typically characterized by the contact resistance ($R_c$). A small $R_c$ is desired for high ON state current and large gain in amplifiers and photodetectors. The issue of electrical contacts is especially critical in 2D materials for the following four reasons. First, there is a fundamental limit to $R_c$ from a 3D metal to a 2D semiconductor based on the number of conduction pathways: $R_c = h/(2e^2 k_F) = 0.026/(n_{2D})^{0.5} \approx 30 \ \Omega$, where $k_F$ is the Fermi wavevector.[45] Second, due to their exceptionally high surface area to volume ratios, the electronic structure of 2D materials can be strongly perturbed by metal contacts.[20,45,113] Third, unlike bulk semiconductors, substitutional doping has been difficult to realize in 2D materials, which would otherwise be used to ensure proper energy level alignment with the metal. Finally, a van der Waals tunnel barrier typically exists between the metal contact and the 2D semiconductor that further increases $R_c$.[45] Following bulk semiconductors, the early efforts in 2D materials were focused on identifying suitable metals to selectively interface with the conduction band (CB) or valence band (VB) for electron or hole transport, respectively. However, deviations from the ideal Schottky-Mott model can occur by Fermi level pinning by surface states (Bardeen model), metal-induced gap states (Heine model), or defect-induced gap states.[40,45] Thus, the actual Schottky barrier height ($\Phi_{SB}$) is often independent of metal work function ($\Phi_M$) or scales linearly as $\Phi_{SB} \sim K_c \Phi_M$ with $K_c$ < 1.

Two-dimensional materials often show unique Fermi-level pinning behavior. Even an ideal Au-MoS$_2$ contact without any defects involves strong interactions between Au and S atoms, which weakens the Mo-S bond of MoS$_2$.[113] Consequently, even in the presence of the intervening S atom, the Mo $d$-orbital is rehybridized with the Au $d$-orbital. Since the Mo $d$-orbital forms 80% of the CB, the Fermi level is pinned in the upper half of the bandgap, explaining the typical n-type behavior independent of contact metal. Beyond the resulting interface gap states, the strong dipole interaction also modifies the local work function of the metal by inducing charge modulation (Fig. 6a, b).[113] Inclusion of defects and processing residues can further complicate the contact behavior, which helps explain seemingly contradictory reports of $\Phi_{SB}$ values. Nevertheless, a systematic study revealed the factor $K_c = 0.1$ for $MoS_2$ contacts with Sc, Ti, Ni, and Pt using a thermionic emission model.[86] The Fermi level can also be de-pinned to reduce $\Phi_{SB}$ by up to ~60% by introducing a tunnel barrier (e.g., $Ta_2O_5$, MgO, hBN, and $TiO_2$).[114] Recently, two additional approaches were reported to significantly improve the contacts with 2D materials. In particular, contacting the 1D edge of graphene from the side, instead of depositing metal on top of the basal plane, strongly enhances in-plane injection of electrons, significantly reducing $R_c$.[115] In addition, the semiconducting 1H phase of $MoS_2$ can be converted to the metallic 1T phase $via$ Li ion intercalation under the metal contacts to achieve $R_c$ as low as 300 Ω-µm (Fig. 6b).[57] However, the generalizability of these approaches is unproven, which suggests that contact engineering of emerging 2D materials will likely remain an active topic of future research.

## 5.4 Doping, defects, and alloys

Much of the success of conventional semiconductors can be attributed to the ability to dope them n-type or p-type without significantly compromising intrinsic properties, while also being able to control bandgap through alloying. In an effort to translate these attributes to 2D materials, significant efforts have been made to dope TMDCs by contact engineering and chemical functionalization.[10,60] As previously discussed, most metals induce Fermi level pinning near the CB in $MoS_2$, resulting in n-type doping. On the other hand, high-purity $MoO_3$ (VB ~ 6.3 eV) contacts have enabled p-type $MoS_2$ transistors.[116] In addition, the all-surface nature of monolayer materials can be exploited for chemical doping. For example, PL measurements have shown p-type and n-type doping in $MoS_2$ following functionalization with 7,7,8,8-tetracyanoquinodimethane (TCNQ) and nicotinamide adenine dinucleotide (NADH), respectively.[117] TCNQ (NADH) increased (decreased) the PL intensity of the neutral exciton in monolayer $MoS_2$ *via* extraction (injection) of excess electrons that favor trion formation. Among other 2D materials, p-type BP has been converted to ambipolar or n-type conduction by using different metal contacts.[12,13]

Structural disorder such as defects and grain boundaries are unavoidable in wafer-scale growth of polycrystalline 2D materials.[118] Single S vacancies, double S vacancies, and 5|7 defects in grain boundaries in TMDCs have shown considerable density of states within the band gap, and thus can act as non-radiative recombination sites for photoexcited carriers that limit quantum yield in optoelectronic devices.[118,119] On other hand, in ambient conditions, S vacancies can stabilize excitons by charge transfer from an adsorbate such as $O_2$ or $N_2$, which increases PL efficiency.[120] Since this type of enhancement mechanism breaks down in the high quantum yield limit, it is more effective to passivate S vacancies when optimizing quantum yield.[121]

In bulk semiconductors, defects are sometimes used to store charge for memory-based devices such as defects at the $SiO_2$–SiN interface being used as an alternative to floating-gate flash memory.[40] Defects in TMDCs have also shown interesting properties that can be exploited for similar applications. In particular, transmission electron microscopy (TEM) has revealed migration of S vacancies within monolayer $MoS_2$ *via* the formation of out-of-plane complexes.[122] DFT calculations also predict anomalous pathways of defect hopping within 2D materials.[123] Namely, a 5|7 defects in S-polar grain boundaries can migrate into the grain by forming a 4|6 defect and a double S vacancy (Fig. 6d). This mechanism is consistent with the observation of single S

vacancies in mechanical exfoliated $MoS_2$ and predominantly double S vacancies and 4|6 defects in CVD-grown $MoS_2$.[118,122] Since defect migration is assisted by grain boundaries, devices fabricated from polycrystalline $MoS_2$ show field-driven defect motion that has been exploited to realize memristive behavior (Fig. 6e).[124] Due to the atomically thin nature of 2D $MoS_2$, this memristive switching can be controlled by an external gate to achieve functions that hold promise for emerging brain-like computing and non-Boolean logic architectures.

Alloying of TMDCs is especially promising because they form one coherent family of compounds with similar structure yet widely different properties. For example, the bandgap of ternary compounds can be varied continuously between two extremes of the constituent binary compounds. Two approaches have been employed by substituting either the transition metal or chalcogen atom.[125-127] Mixing of chalcogens in $MoX_2$ (X = S, Se, or Te) is consistent with the Vegard law (i.e., lattice constant varies linearly with mixing fraction). The free energy of mixing is negative for all combinations, but the $MoS_2$–$MoSe_2$ alloys is expected to be the most stable.[125] In this case, S-Se nearest neighbors are preferred from entropy considerations, which implies that chalcogen mixing does not show a tendency for long-range order or segregation. Indeed, the optical bandgap of CVD-grown $MoS_{(2-x)}Se_x$ has been found to vary linearly between 1.85 eV and 1.6 eV as the fraction of Se is increased from 0 to 2 (Fig. 6f).[126] Similarly, stable alloys of $Mo_{(1-x)}W_xS_2$ have been reported with the bandgap changing parabolically with W composition, which is consistent with DFT calculations.[125,127] Although random substitution in alloys breaks down crystalline translational symmetry, band delocalization is still preserved in TMDCs due to significant $d$-character at band extrema. Interestingly, uneven distribution of chalcogens in top and bottom layers (Fig. 1a) can break down mirror symmetry in the z-direction. In the extreme limit,

Janus monolayers have been grown by CVD where S and Se occupy bottom and top layers, respectively.[128]

Doping of other 2D materials, such as BP with N, has been proposed but experimental demonstrations are scarce. Ternary compounds of PMTCs also show wide tunability of bandgap such as $GaS_{(1-x)}Se_x$ showing PL peak shifts between 2.5 eV and 2.0 eV as the Se composition is varied from 0 to 1.[64] Alloying between (In, Ga) and (S, Se, Te) is a fertile topic of exploration due to the high mobility and photoresponsivity of these compounds. Finally, TMDC alloys involving transition metals from different groups have been explored in the bulk,[24] but comparable monolayer alloys have not yet been achieved.

## 6. Van der Waals heterojunctions

One of the most exciting prospects for 2D materials is the ability to stack them in atomically abrupt heterojunctions bonded solely by van der Waals (vdW) interactions. Unlike covalently bonded heterojunctions in bulk semiconductors, vdW heterojunctions avoid dangling bonds and trapped charges. This relaxation of the lattice matching condition allows vast permutations of 2D material heterojunctions.[33,36] Graphene was the first 2D material used in vdW heterojunctions as a barrister with bulk Si and as an FET with hBN.[31,37] Most graphene-based vdW heterojunctions rely on tuning the graphene Fermi level by electrostatic doping on nearly defect-free hBN substrates. However, the zero bandgap of graphene is incompatible with many semiconductor heterojunction applications such as light emitters and solar cells. On the other hand, vdW heterojunctions from 2D semiconductors have been used in a range of advanced devices including Esaki diodes, tunnel transistors, gate-tunable photovoltaic cells, light emitting diodes (LEDs), anti-ambipolar rectifiers, and photodetectors.[34,35,39,129-132] Instead of surveying individual devices, here, we focus instead on the broader concepts of band alignment, device architecture, and bottleneck issues.

Most 2D semiconducting vdW heterojunctions can be divided into type-II (staggered) or type-III categories (broken) (Fig. 7a, b). Type-I (straddling) heterojunctions are mostly used to create a 2D electron gas in bulk semiconductors and are less relevant here.[40] The most significant materials parameters for determining heterojunction properties, such as CB and VB positions, electron *versus* hole doping, and direct *versus* indirect band gap are listed in Fig. 2. Versatile type-II heterojunctions provide rectification for signal amplification, charge transfer for photon energy conversion, and charge recombination for light emission.[40] Due to the weak screening in 2D materials, the Fermi level at the heterojunction can be tuned by the gate electrode, which allows for field-driven control of rectification and photoresponse in p-n heterojunction diodes. For example, the short-circuit current in the type-II heterojunction $MoS_2/WSe_2$ varies with gate bias (Fig. 7a).[46,133] Varying the bandgap with thickness and integrating dual gates present additional degrees of freedom for further control of charge transport. For example, dual-gated few-layer $MoS_2/WSe_2$ heterojunctions have achieved both band-to-band tunneling in reverse bias (Zener diode) and negative differential resistance in forward bias (Esaki diode).[131]

Another promising application of vdW heterojunctions are low-power tunneling devices that achieve switching below the thermal limit of ~60 mV/decade in FETs. In addition, room temperature negative differential resistance has been realized by type-III $BP/SnS_2$ vdW heterojunctions with a peak-to-valley ratio of 1.8 (Fig. 7b). A quick look at Fig. 2 suggests other possible combinations of materials for type-III heterojunctions such as monolayers of p-type SnS and n-type $HfS_2$ that could yield higher peak-to-valley ratios. It should be noted that this negative differential resistance behavior is akin to a bulk Esaki diode and qualitatively different from negative differential resistance switching in graphene-based vdW heterojunctions that rely on forbidden tunneling of electrons that cannot satisfy momentum conversation at higher Fermi levels, a condition uniquely enabled by the linear energy dispersion of graphene.

Van der Waals heterojunctions can be fabricated in two geometries, namely lateral and vertical (Fig. 7c).[134-136] In the lateral geometry, two semiconductors are stacked in a misaligned geometry so that electrodes can be placed on non-overlapping areas. In contrast, the vertical geometry uses electrodes in direct contact with the two sides of the heterojunction (Fig. 7c). While the lateral geometry allows better control of electrostatics, the current density is limited by the series resistance of non-overlapping areas.[38,39,135,136] Atomically thin lateral heterojunctions have also been realized by stitching p-type and n-type $MoS_2/WSe_2$ semiconductors by CVD growth (Fig. 7c).[130] On the other hand, the vertical geometry allows larger current density at the expense of electrostatic control and leakage current from intrinsic crystal defects in 2D materials.[133]

Integrating 2D materials with materials of different dimensionality such as 0D quantum dots (QDs), 1D nanotubes and nanowires, and 3D bulk semiconductors expands the phase space of possible device functions.[5,38,39,135,137,138] For example, record high photoresponsivity ($\sim10^7$ A/W) has been achieved in 2D-0D graphene-PbS QD heterojunctions by taking advantage of the high mobility of graphene and the high optical absorption of QDs.[46] On other hand, Förster-like nonradiative energy transfer was observed between $MoS_2$ and core-shell CdSe/CdZnS QDs.[138] Tunnel transistors using 2D-3D $MoS_2$-germanium heterojunctions achieved a subthreshold swing of $\sim32$ mV/decade and net operating voltage $\sim0.1$ V.[132] Similarly, 2D-1D $MoS_2$-CNT and 1D-3D CNT-IGZO vdW heterojunctions enable anti-ambipolar characteristics that have the potential to simplify circuits used in signal processing.[39,137] Integrating 2D materials with organic semiconductors presents additional opportunities for photovoltaics. For example, heterojunctions between $MoS_2$ and organic donor polymers have demonstrated internal quantum efficiencies up to 40% in a ~20 nm thick solar cell.[41] Pentacene-$MoS_2$ vdW heterojunctions have further shown ultrafast hole transfer (~6 ps) and long-lived charge separated states (~5 ns, which is 2–80 times longer than 2D-2D heterojunctions) (Fig. 7d).[119]

## 7. Composite films

The large surface area to volume ratio of nanomaterials has been extensively exploited for chemical sensing, catalysis, electrochemistry, and energy storage. 2D materials provide the ultimate scaling in one dimension while preserving robust mechanical strength in the orthogonal plane. The van der Waals interaction between layers also enables facile means of exfoliation by intercalation of alkali metal ions or mechanical shear forces.[7,48] These factors have naturally led to the exploration of composite films of 2D materials. Since methods for solution exfoliation and thin-film assembly have been reviewed elsewhere,[7,48,139,140] we focus here on issues related to electronic transport in composite films.

Fig. 7e, f show the typical morphology of a composite film where 2D flakes (area < 1 μm$^2$) are assembled in random orientation within the plane while maintaining a certain degree of anisotropy out-of-plane.[140,141] Charge transport within an individual flake can be diffusive (band-like) or hopping-type depending on several factors such as intrinsic electronic structure and delocalization of defect states.[93,96,98] In the absence of post-fabrication sintering, inter-flake transport is expected to be variable-range hopping with the probability of hopping scaling as ~exp[–2R/α–W/kT], where R, α, and W are the distance between nearest neighbor flakes, attenuation length, and energy difference between conduction states in the flakes, respectively (Fig. 7f). Thus,

compact films with flakes of uniform size and thickness are preferred for higher conductivity. For example, density gradient ultracentrifugation of $ReS_2$ flakes enabled by weak interlayer coupling resulted in electrical conductivity that is ~$10^8$ times higher than $MoS_2$ films made from standard sonication.[142] Since some 2D materials such as BP tend to degrade in ambient conditions, solution-processing in anhydrous solvents allows a method for preserving the intrinsic properties of individual flakes.[49,143] Ultimately, highly conducting films are desired that are thin enough for electrostatic control in a field-effect geometry. So far, this goal has been elusive, although gating was recently achieved by ion-gel penetrating composite dielectrics made from hBN flakes (Fig. 7e).[141] Since mechanically exfoliated TMDCs show exceptionally high photoresponsivities (e.g., ~$10^5$ A/W for $ReS_2$),[35] a more promising application of 2D material composites are photodetectors where they potentially compare favorably with commercial Si photodiodes (~1 A/W).[40]

## 8. Future outlook

The initial exploration of post-graphene 2D materials followed the footsteps of graphene research, resulting in a rapid demonstration of a wide range of electronic devices and emergent charge transport phenomena. At present, while the fundamental understanding of electronic transport in individual 2D materials is reaching maturation, several bottleneck issues remain for wafer-scale practical applications. From the materials science perspective, significant challenges surround the realization of robust wafer-scale growth methods that yield uniform and controllable thickness. For example, early research in powder-vaporization (or CVD) resulted in small-area films with insufficient thickness control.[43,47,118] Molecular beam epitaxy (MBE) and atomic-layer deposition (ALD) also did not fully resolve these issues. Recently, metal-organic chemical vapor deposition (MOCVD) has produced monolayers of $MoS_2$ and $WS_2$ over 4-inch wafers with respectable mobilities of ~30 cm$^2$/Vs (Fig. 8a), although the reported growth time of 26 hours is uncomfortably long for most industrial processes.[30] Controlled substitutional doping is another major objective to enable tunable bandgaps and contact engineering (Fig. 8b). Towards this end, ternary TMDC compounds have shown initial success, but quaternary compounds such as $M_xW_{(1-x)}S_ySe_{(2-y)}$ and $In_xGa_{(1-x)}S_ySe_{(1-y)}$ are relatively uncommon. Furthermore, due to the limited understanding about the effects of doping on electronic transport in two-dimensions, more readily accessible synthetic materials would facilitate this direction of investigation. Finally, some of the most promising 2D materials such as elemental monolayers of BP and silicene are unstable in ambient conditions.[14-16,73,144] Therefore, robust passivation schemes are needed to retain their high performance in practical environments, and effective processing strategies are required for seamless integration into circuits (Fig. 8c).[44] Since electronic transport is more sensitive to degradation-induced traps and defects than common spectroscopy tools, comparative assessment of different passivation schemes should employ tools beyond standard metrology.[144]

The chemistry of 2D materials also presents unique challenges and opportunities compared to conventional bulk semiconductors. The high surface area to volume ratio of 2D materials implies that chemical functionalization can dramatically affect properties with the lone electron pair on the chalcogen atoms in TMDCs and phosphorus atoms in BP providing opportunities for Lewis acid/base chemistry and radical-based chemistry.[49,145] The challenge, however, is to identify covalent functionalization chemistries that enable electronic structure tuning without perturbing band delocalization and charge carrier mobility (Fig. 8d). In terms of optoelectronics, the conversion efficiencies of photons to electrons in photodetectors and solar cells, and electrons to photons in light-emitting diodes are ultimately dictated by the relative timescales of radiative and non-radiative processes.[74,99] Therefore, chemical routes for passivating non-radiative recombination sites hold significant promise for enhancing optoelectronic technologies based on 2D materials. Mixed-dimensional vdW heterojunctions that mate 2D materials with 0D or 1D nanostructures also present a broad and relatively unexplored phase space for fundamental research and application development (Fig. 8f).[138]

From the device engineering perspective, future efforts are likely to explore the ultimate scaling limit for sub-5 nanometer transistors, particularly because the high mobility and immunity to short-channel effects in Group VI TMDCs meet the requirements set by ITRS.[21] One can further imagine a scenario where all components of electronic devices, namely semiconducting channels, metallic contacts, and insulating dielectrics, are composed of 2D materials in a seamless fashion (Fig. 8g).[146] Thus far, a limited number of high-quality 2D material heterojunctions have been grown, which implies that significant future research is still required to achieve scalable all-2D electronic architectures. In addition, the extrinsic effects of metal contacts and interfacial disorder in 2D materials remain a major challenge for 2D device engineering (Fig. 8h). Side-contact geometries, 1T phase contacts, and hBN substrates have shown promise in individual devices, but a scalable route for uniformly realizing these device concepts at the wafer-scale has not yet been established. We again stress that conventional techniques of contact-engineering usually fail in 2D materials, and thus a fundamental understanding of charge injection and collection is critical for identifying unconventional solutions. Similarly, despite tremendous activity in vdW heterojunctions at the single device level, the scalability of vdW heterojunctions remains an outstanding issue. The challenge in this case is two-fold. Not only do individual components need to be uniform at the wafer-scale with atomic-scale thickness control, but the different layers should also be aligned with the precision of the node-size (Fig. 8i). Drawing lessons from the long success of self-aligned gates in conventional Si FETs, analogous self-alignment methods are likely to yield breakthroughs in vdW heterojunctions. Finally, although this article primarily focused on electronic transport in individual 2D materials, more complex materials systems such as partially oxidized chalcogenides (e.g., $Bi_2O_2Se$),[147] naturally occurring complex heterojunctions (e.g., franckeite and alternative layers of PbS and $SnS_2$),[148] and emerging synthetic vdW heterojunctions present vast promise for future charge transport experiments.

## 9. Summary points

1. The vast library of 2D layered materials show diverse electronic structure and properties that are strongly dependent on thickness and external fields.

2. Theoretical and experimental data on scattering mechanisms show close agreement and reveal that intrinsic mobility limits are realized with tight control of interfacial homogeneity as has been demonstrated with hexagonal boron nitride.

3. Extrinsic effects such as contacts, defects, substrate effects, and environment limit performance in most electronic devices, thus necessitating the development of mitigating surface and interfacial treatments.

4. Group VI transition metal dichalcogenides, black phosphorus, and indium selenide show significant promise for ultra-thin body electronics and optoelectronics.

5. Stacking of 2D layers in van der Waals heterostructures show emergent phenomena and novel device characteristics that can be further generalized and enhanced by employing additional low-dimensional nanostructures.

6. Uniform large-area growth, controlled stitching of different 2D materials, and alloying schemes are among the most significant challenges for practical realization of 2D electronics.

**Disclosure Statement**

The authors are not aware of any affiliations, memberships, funding, or financial holdings that might be perceived as affecting the objectivity of this review.

**Acknowledgements**

The authors acknowledge support from the National Science Foundation Materials Research Science and Engineering Center (DMR-1121262) and 2-DARE program (EFRI-1433510).

**Figures**

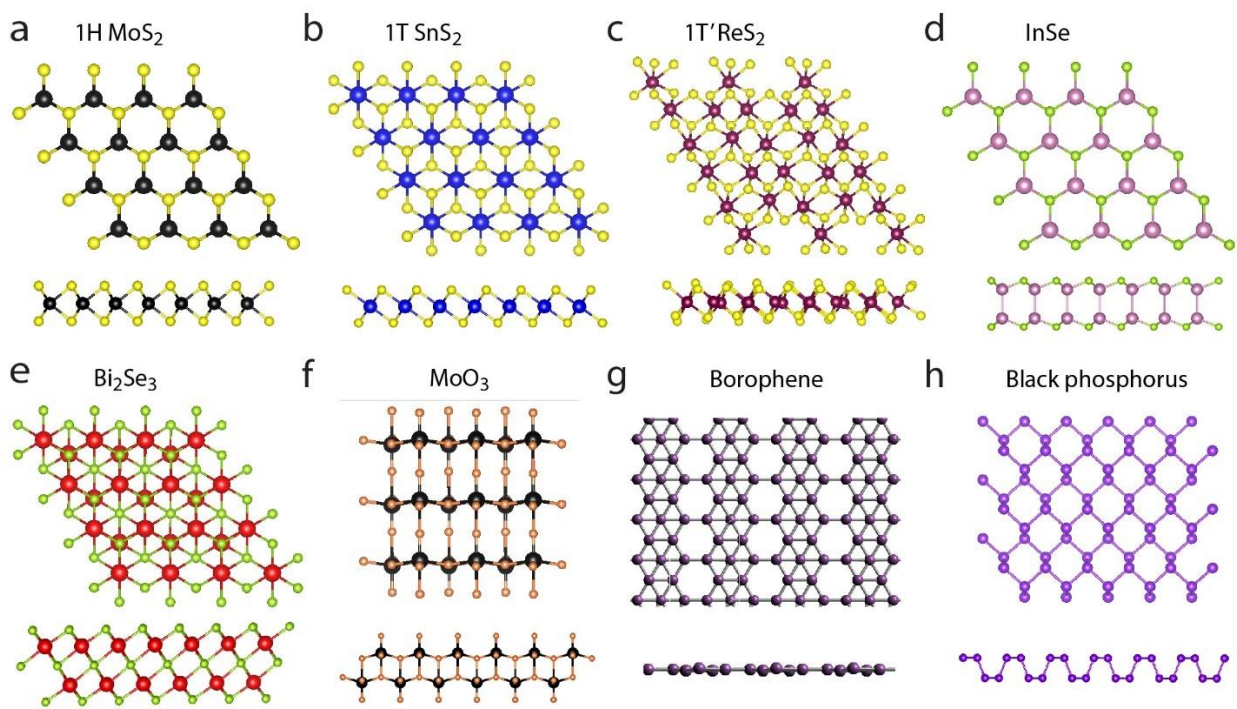

**Figure 1. Crystal structure of post-graphene 2D materials (top and side views).** (a) Trigonal prismatic crystal structure of 1H TMDCs such as MoS₂ and WS₂. (b) Octahederal crystal structure of 1T TMDCs such as WTe₂ and PTMCs such as SnS₂. (c) Distorted octahedral crystal structure in 1T' form of TMDCs such as ReS₂ and SiS₂. (d) Crystal structure of III-VI monochalcogenides such as InSe and GaS. (e) Crystal structure of monolayer Bi₂Se₃. (f) Crystal structure of monolayer MoO₃. (g) Crystal structure of one of the phases of monolayer borophene grown on Ag (111). (h) Puckered anisotropic crystal structure of monolayer black phosphorus (BP).

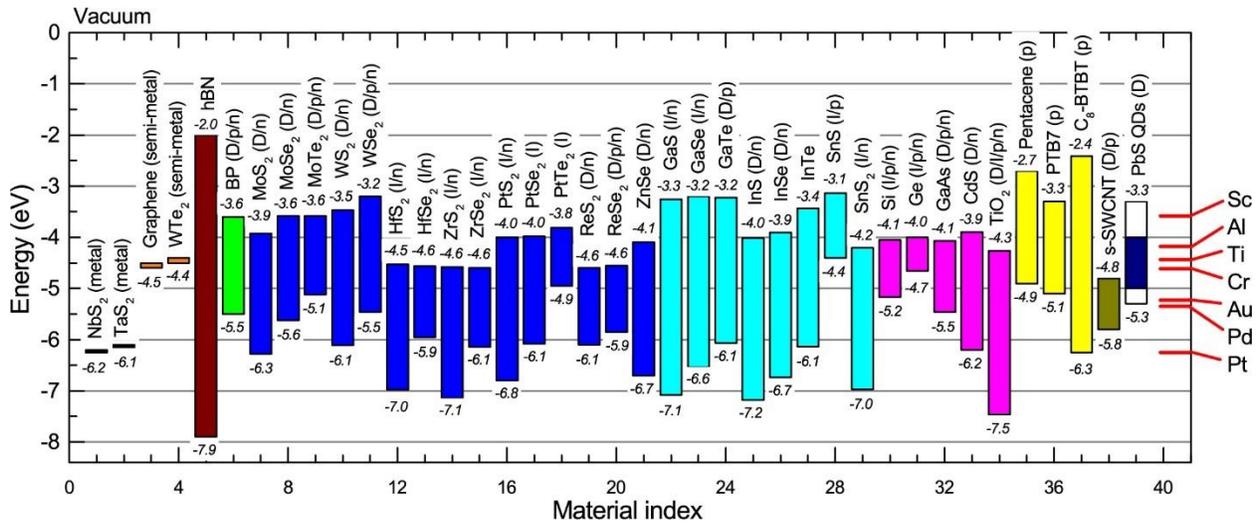

**Figure 2. Electronic band parameters of selected 2D materials.** Position of conduction band (CB) minima and valence band (VB) maxima for selected 2D monolayers and conventional semiconductors with respect to vacuum. The work function is provided for metals and semimetals. Materials are classified by color as: 2D metals (index 1,2, black); 2D semi-metals (3,4, orange), hBN (5, brown); BP (6, green); 1H TMDCs (7-21, blue); PTMCs (22-29, cyan); inorganic semiconductors (30-34, magenta); organic semiconductors (35-37, yellow); semiconducting single-walled carbon nanotubes (s-SWCNTs) (38, dark yellow); PbS quantum dots (39, dark blue). The red lines on the right axis show the work functions of common metal contacts. Information in parentheses: D = direct band gap; I = indirect band gap, p(n) = predominantly hole (electron) conduction; p/n = both electron and hole conduction or ambipolar

conduction; D/I = both direct and indirect gap possible depending on phase. In case the electron affinity (CB) is not available for monolayers, this value is taken from the bulk. In case of large differences in optical and electronic bandgaps (e.g., 1.8 eV *versus* 2.4 eV for MoS₂, respectively), electronic bandgaps are plotted. All materials parameters are collected from the following references.[6,8,17,18,24,40,41,64,65,75,104,149-151]

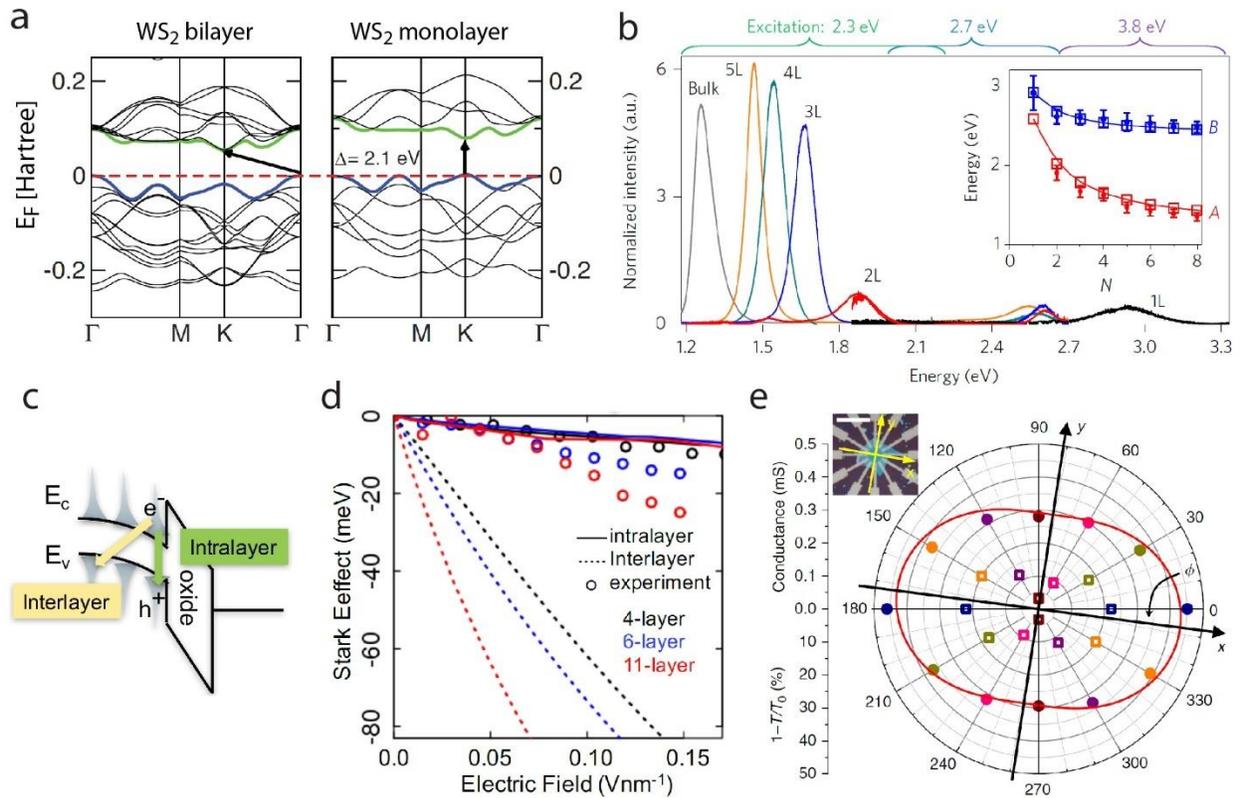

**Figure 3. Tuning electronic structure with thickness and field.** (a) Band structure of monolayer and bilayer WS₂ calculated by density functional theory (DFT). A direct transition at the K point for monolayers and an indirect transition (K-Γ) for bilayers are shown by arrows. VB and CB extrema are highlighted. (b) Photoluminescence spectra of InSe flakes with varying thicknesses measured with three excitation energies of 2.3, 2.7, and 3.8 eV. All photoluminescence (PL) peaks are normalized with number of layers (N) except the bulk (~45 nm). Inset shows N-dependent PL peaks for excitons A and B with DFT calculations (squares). (c) Schematic of interlayer and intralayer excitonic transitions in a few-layer TMDC field-effect

transistor (FET). (d) Bandgap change at K points with external field for 4, 6, and 11 layers of $MoS_2$. Open circles show the experimental PL peak shift. Solid and dashed lines show the calculated peak shift for intralayer and interlayer excitons, respectively. (e) Polar plot showing in-plane anisotropy in electrical conductivity and optical absorbance of black phosphorus (BP). Angle-resolved conductivity and polarization-resolved infrared (2700 $cm^{-1}$) relative extinction was measured along six directions on the BP flake (inset). Scale bar: 50 μm. Figure panels are adapted with permission as follows: panel *a*, Reference[25], copyright @ 2011 American Physical Society; panel *b*, Reference[17], copyright @ 2017, Nature Publishing Group (NPG); panels *c* and *d*, Reference[27], copyright @ 2017, arxiv.org; panel *e*, Reference[83], copyright @ 2014, NPG.

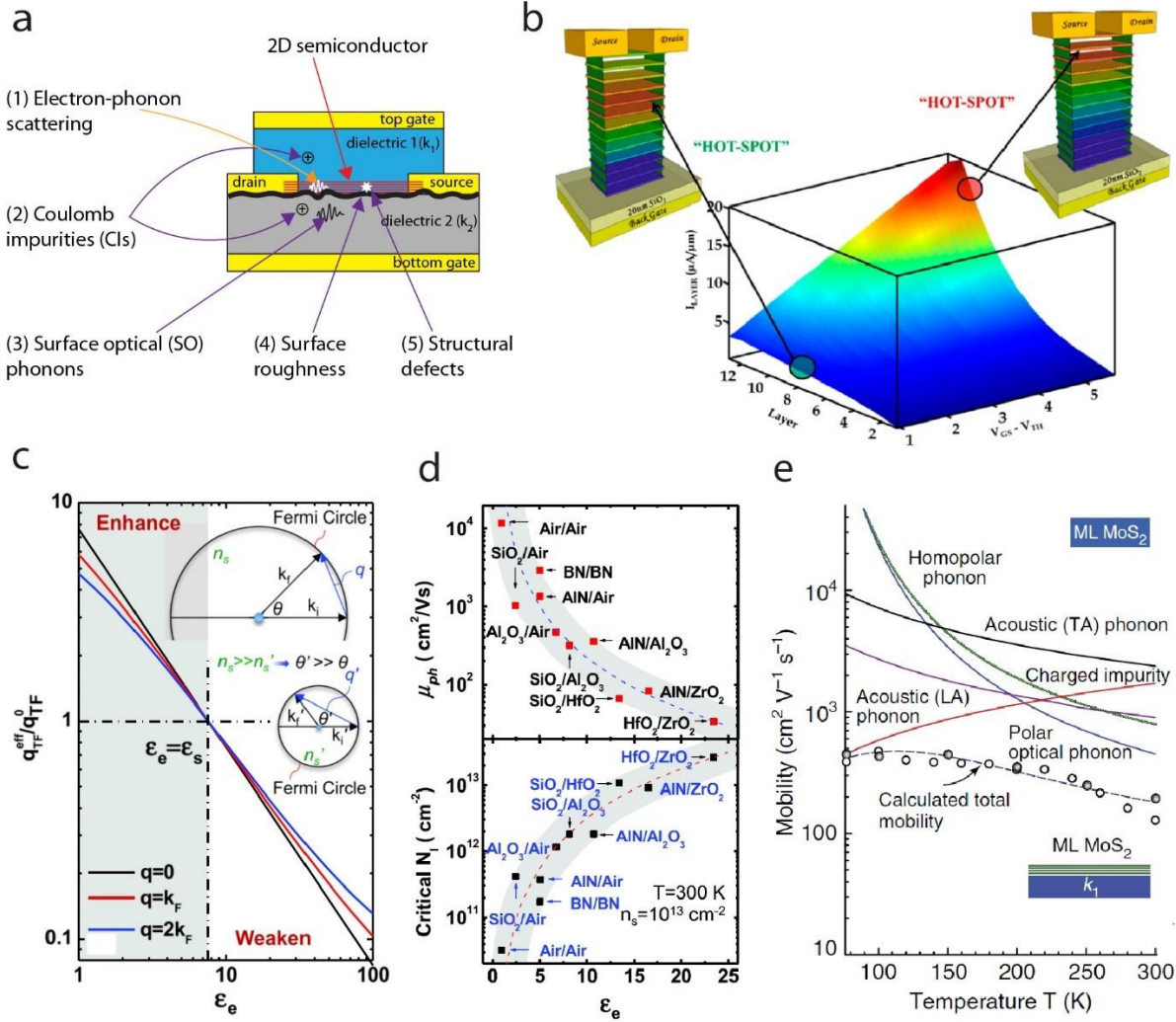

**Figure 4. Charge screening and scattering in 2D materials.** (a) Schematic of a 2D layered FET in a dual-gated geometry. $k_1$ and $k_2$ are dielectric constants for dielectric 1 and 2, respectively. (b) Schematic of current distribution in a 13-layer $MoS_2$ FET showing position of the maximum current density (hot spot) determined by competing effects of conductivity and charge screening. (c) Effect of dielectric mismatch on screening in monolayer $MoS_2$. The inset shows scattering of conduction electrons on the Fermi circle from wavevector $\mathbf{k_i}$ to $\mathbf{k_f}$ by a phonon of wavevector $\mathbf{q} = \mathbf{k_f} - \mathbf{k_i}$. The Fermi circle is larger for larger carrier density ($n_s$), which implies that the scattering angle ($\theta$) decreases with increasing $n_s$. (d) Top: Calculated values of phonon-limited mobility of monolayer $MoS_2$ for different combinations of dielectrics. $\varepsilon_e = (k_1+k_2)/2$ is the average dielectric constant seen by $MoS_2$. Phonon-limited mobility values decrease with increasing $\varepsilon_e$ due to SO phonon scattering. Bottom: Critical density of charged impurities (CIs) $N_I$ at which phonon-limited mobility equates to CI-limited mobility. High-$k$ dielectric is more effective when $MoS_2$ is already disordered with CIs, resulting in marginal improvements in mobility. (e) Experimental mobility values (circles) compared with mobilities calculated from different scattering mechanisms (lines) for a few-layer $MoS_2$ FET on a $SiO_2$ substrate. Panels are adapted with permission as follows: panel $b$, Reference[85], copyright @ 2013, American Chemical Society; panels $c$ and $d$, Reference[87], copyright @ 2014, American Physical Society; panel $e$, Reference[95], copyright @ 2012, Nature Publishing Group.

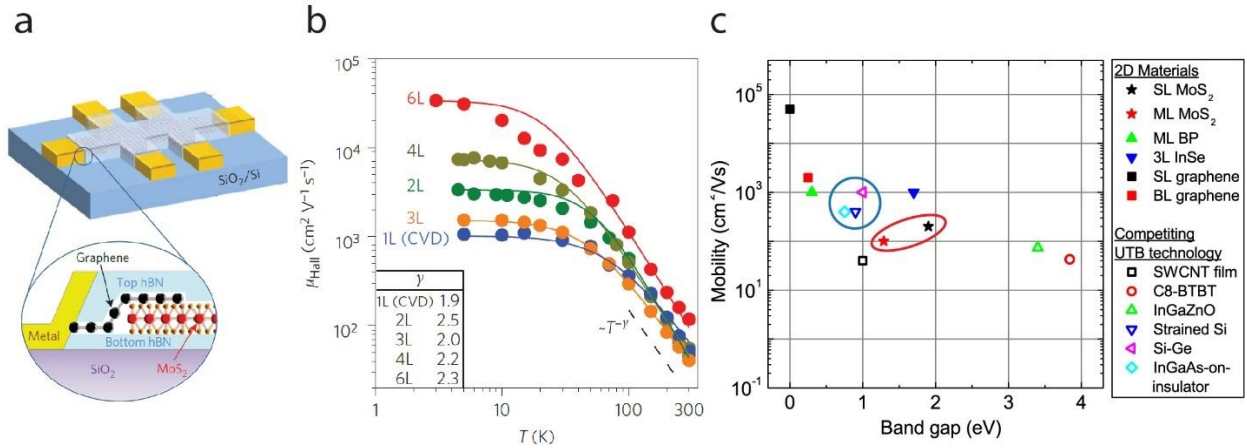

**Figure 5. Field-effect mobility and performance metrics.** (a) Schematic of a monolayer $MoS_2$ FET sandwiched between hBN layers and contacted by graphene. (b) Temperature-dependent mobility for different thicknesses of $MoS_2$. Adapted with permission from Reference[29], copyright @ 2015, Nature Publishing Group. (c) Mobility and bandgap of the most promising 2D semiconductors compared with competing ultra-thin body (UTB) technologies. Mobility values of monolayer $MoS_2$, few-layer $MoS_2$, few-layer BP, trilayer (3L) InSe, and graphene are taken from ref[29], ref[95], ref[12], ref[17], ref[115], and ref[28], respectively. Mobility data for SWCNT thin films, $C_8$-BTBT, and inorganic metal oxide InGaZnO are taken from ref[106], ref[104], and ref[105], respectively. Mobility values of stained Si, Si-Ge alloys, and InGaAs-on-insulator are taken from ref[101], ref[102], and ref[103], respectively. Blue and red ovals show mobility range of commercial UTB and TMDCs, respectively.

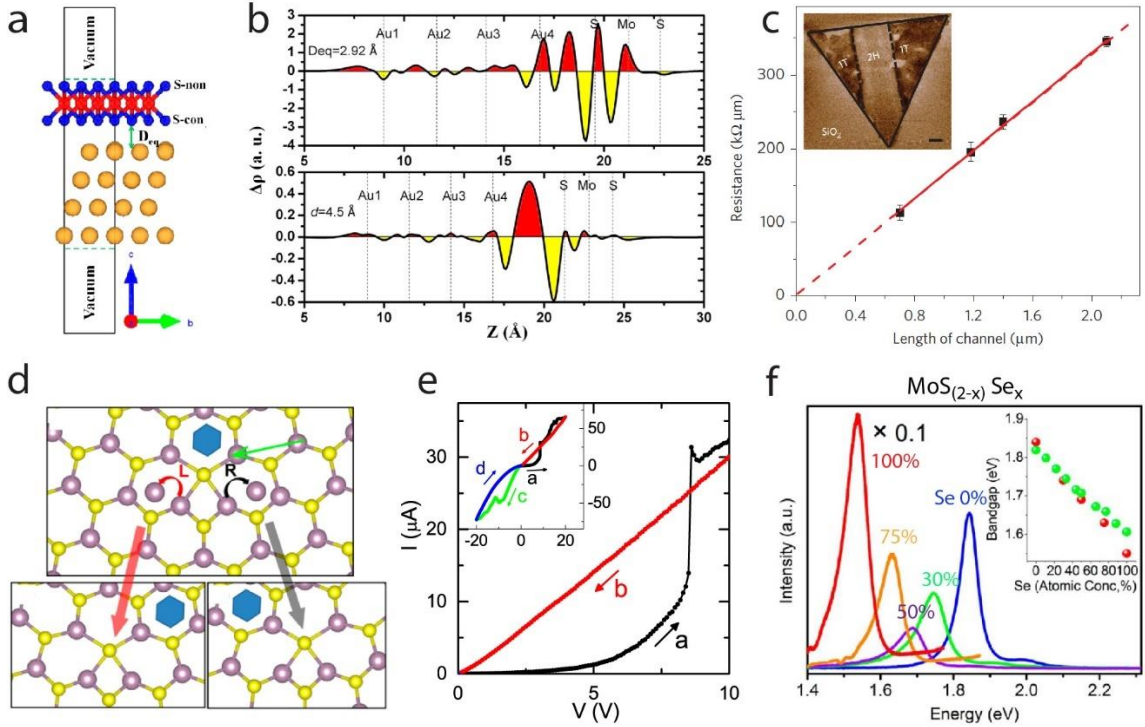

**Figure 6. Role of contacts, defects, and alloying.** (a) Schematic of atomic-level DFT calculations for a contact between Au and monolayer $MoS_2$. (b) Calculated electron density showing cycles of charge accumulation and depletion from Fermi level pinning and orbital-mixing between S and Au atoms. (c) Plot of resistance with channel length $L_{ch}$ in a FET with semiconducting 2H $MoS_2$ as the channel and metallic 1T $MoS_2$ as the electrodes. Extrapolation to $L_{ch} = 0$ gives a contact resistance $R_c \sim 300\ \Omega$-µm. The inset shows an atomic force microscopy (AFM) image of the 2H and 1T regions in a $MoS_2$ flake. (d) DFT calculations showing the anomalous mechanism of migration of a 6|4 defect complex in monolayer $MoS_2$ assisted by an intercalated transition metal atom. (e) Memristive current-voltage (*I-V*) characteristics of a device based on CVD-grown $MoS_2$ with grain boundaries. (f) PL peak shift in a CVD-grown alloy $MoS_{(2-x)}Se_x$ with varying content percentage of Se. Inset: Optical bandgap *versus* Se concentration (%). Panels are adapted with permission as follows: panels *a* and *b*, Reference[113], copyright @ 2014, American Chemical Society (ACS); panel *c*, Reference[57], copyright @ 2014, Nature Publishing Group (NPG); panel *d*, Reference[123], copyright @ 2015, ACS; panel *e*, Reference[124], copyright @ 2015, NPG; panel *f*, Reference[126], copyright @ 2013, ACS.

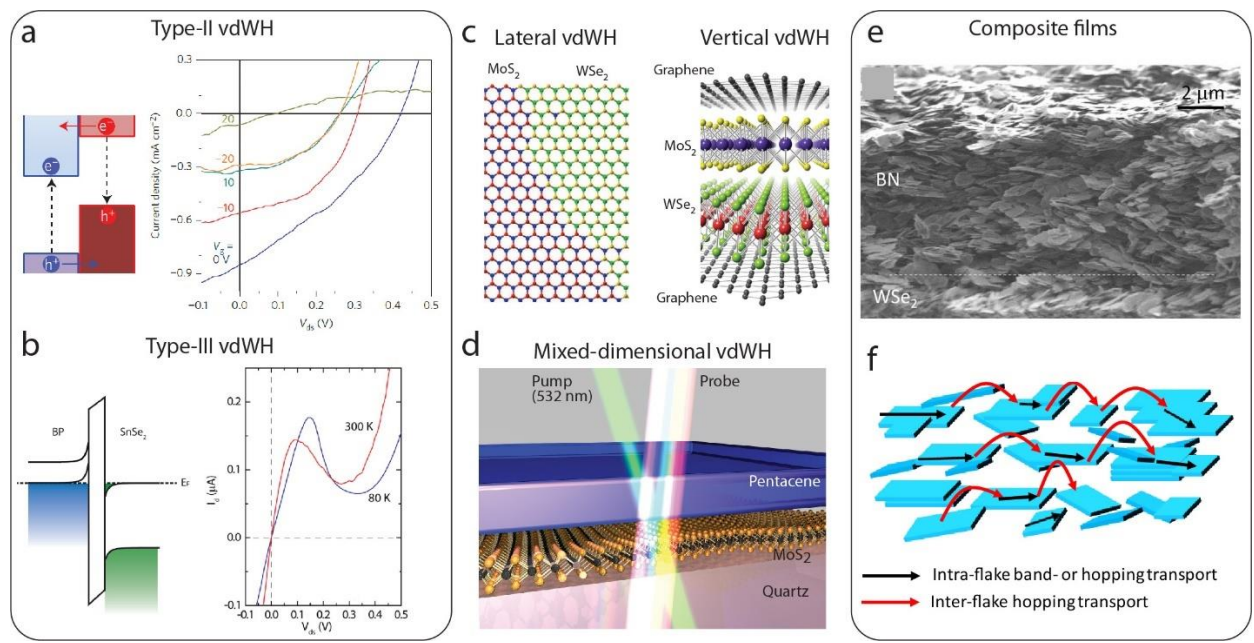

**Figure 7. Van der Waals heterojunctions and composite films.** (a) Schematic band diagram and I-V characteristics of a type-II p-n heterojunction diode between $MoS_2$ and $WSe_2$. Short-circuit current is tuned with gate voltage ($V_g$). (b) Schematic of band-diagram and I-V characteristics of a type-III p-n heterojunction diode between BP and $SnSe_2$ showing negative differential resistance at room temperature. (c) Left: Schematic of a lateral vdW heterojunction between covalently stitched $MoS_2$-$WSe_2$ grown by CVD. Right: Schematic of a vertical vdW heterojunction between $MoS_2$-$WSe_2$ sandwiched between two layers of graphene. (d) Schematic of a mixed-dimensional vdW heterojunction between $MoS_2$ and pentacene probed by transient absorption spectroscopy. (e) Scanning electron microscope image showing morphology of composite bilayer film obtained by sequential printing of $WSe_2$ and BN inks. (f) Schematic of charge transport in a composite film, showing hopping and diffusive transport mechanisms. Panels are adapted with permission as follows: panels *a*, Reference[133], copyright @ 2014, Nature Publishing Group (NPG); panel *b*, Reference[129], copyright @ 2015, American Chemical Society (ACS); panel *c* (lateral), Reference[130], copyright @ 2015, American Association for the Advancement of Science (AAAS); panel *c* (vertical), Reference[133], copyright @ 2014, NPG; panel *d*, Reference[119], copyright @ 2016, ACS; panel *e*, Reference[141], copyright @ 2017, AAAS.

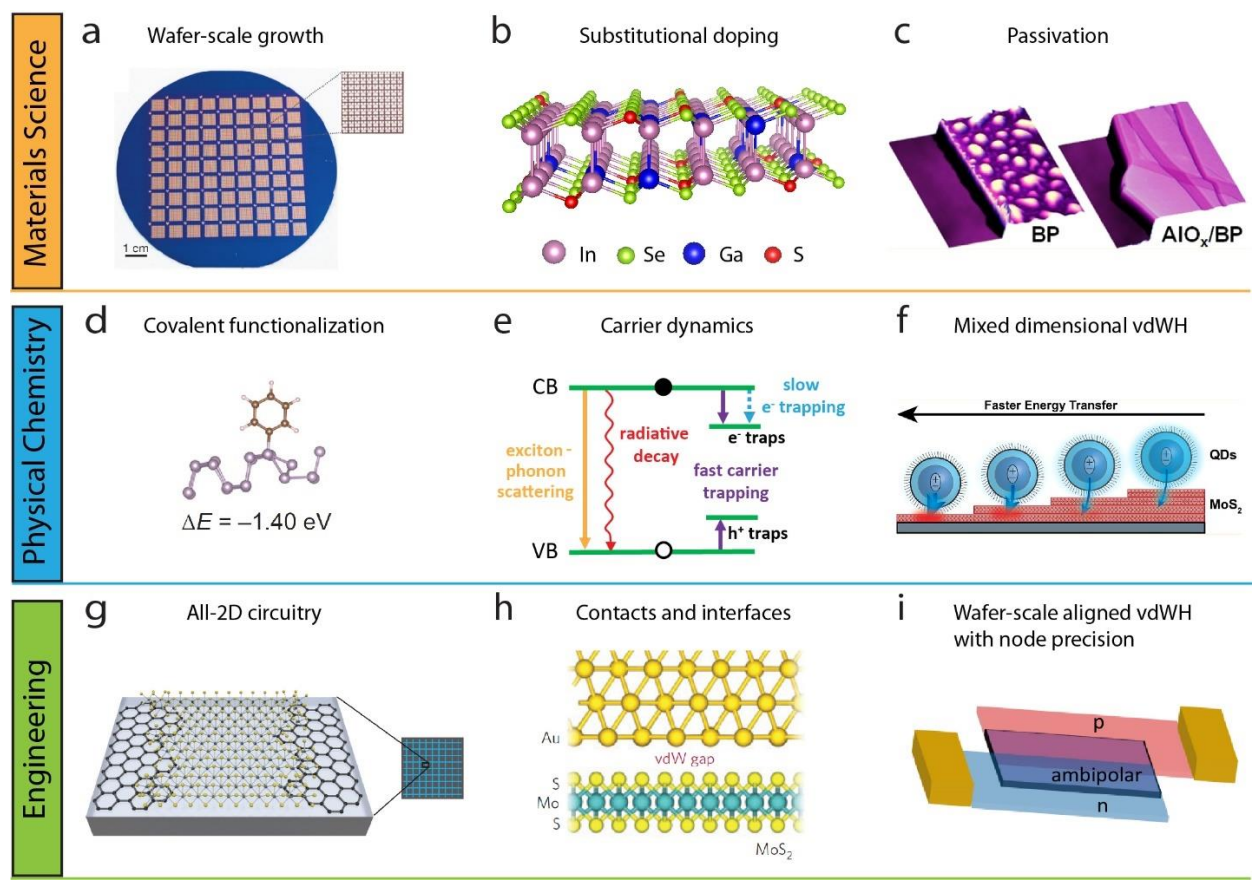

**Figure 8. Challenges and opportunities for electronic transport in 2D materials.** (a) Optical micrograph of a 4-inch wafer with monolayer $MoS_2$ grown by MOCVD. (b) Schematic of InSe alloyed with Ga and S atoms as a potential pathway to substitutional doping. (c) AFM images of unencapsulated BP and a BP flake encapsulated by an ALD-grown $AlO_x$ passivation layer. Similar passivation strategies are desired for reactive 2D materials such as silicene and borophene. (d) Schematic of BP covalently functionalized with an aryl radical. Covalent functionalization holds promise for tailoring bandstructure and acting as a passivation layer. (e) Schematic showing various recombination pathways of photoexcited carriers in $MoS_2$. Minimizing non-radiative pathways is important for achieving high quantum yields in light-emitting diodes and solar cells. (f) Schematic showing energy transfer between a 0D-2D QD-$MoS_2$ heterojunction. (g) Schematic showing a CVD-grown 2D electronic circuit with seamless connection between graphene (contact) and $MoS_2$ (semiconductor). (h) Schematic showing the vdW gap between a metal and a 2D material that acts as tunnel barrier. Effective contact engineering remains a key issue in 2D electronics. (i) Schematic showing wafer-scale vdW heterojunctions between 2D materials. A breakthrough in integration is needed to move beyond currently used transfer methods for micron-sized individual flakes. Panels are adapted with permission as follows: panel *a*, Reference[30], copyright @ 2015, Nature Publishing Group (NPG); panel *c*, Reference[44], copyright @ 2014, American Chemical Society (ACS); panel *d*, Reference[145], copyright @ 2016, NPG; panel *e*, Reference[119], copyright @ 2016, ACS; panel *f*, Reference[138], copyright @ 2014, ACS; panel *g*, Reference[146], copyright @ 2016, NPG; panel *h*, Reference[45], copyright @ 2015, NPG.

# References

(1)    Novoselov, K.; Jiang, D.; Schedin, F.; Booth, T.; Khotkevich, V.; Morozov, S.; Geim, A. Two-Dimensional Atomic Crystals. *Proceedings of the National Academy of Sciences of the United States of America* **2005**, *102*, 10451-10453.

(2)    RadisavljevicB; RadenovicA; BrivioJ; GiacomettiV; KisA. Single-layer MoS2 transistors. *Nat Nano* **2011**, *6*, 147-150.

(3)    Mak, K. F.; Lee, C.; Hone, J.; Shan, J.; Heinz, T. F. Atomically Thin MoS2: A New Direct-Gap Semiconductor. *Phys Rev Lett* **2010**, *105*, 136805.

(4)    Jariwala, D.; Sangwan, V. K.; Lauhon, L. J.; Marks, T. J.; Hersam, M. C. Carbon Nanomaterials for electronics, optoelectronics, photovoltaics and sensing. *Chemical Society Reviews* **2013**, *42*, 37.

(5)    Jariwala, D.; Sangwan, V. K.; Lauhon, L. J.; Marks, T. J.; Hersam, M. C. Emerging device applications for semiconducting two-dimensional transition metal dichalcogenides. *ACS nano* **2014**, *8*, 1102-1120.

(6)    Wang, Q. H.; Kalantar-Zadeh, K.; Kis, A.; Coleman, J. N.; Strano, M. S. Electronics and optoelectronics of two-dimensional transition metal dichalcogenides. *Nature nanotechnology* **2012**, *7*, 699-712.

(7)    Nicolosi, V.; Chhowalla, M.; Kanatzidis, M. G.; Strano, M. S.; Coleman, J. N. Liquid exfoliation of layered materials. *Science* **2013**, *340*, 1420.

(8)    Xu, M.; Liang, T.; Shi, M.; Chen, H. Graphene-Like Two-Dimensional Materials. *Chemical Reviews* **2013**, *113*, 3766-3798.

(9)    Butler, S. Z.; Hollen, S. M.; Cao, L.; Cui, Y.; Gupta, J. A.; Gutiérrez, H. R.; Heinz, T. F.; Hong, S. S.; Huang, J.; Ismach, A. F.; Johnston-Halperin, E.; Kuno, M.; Plashnitsa, V. V.; Robinson, R. D.; Ruoff, R. S.; Salahuddin, S.; Shan, J.; Shi, L.; Spencer, M. G.; Terrones, M.; Windl, W.; Goldberger, J. E. Progress, Challenges, and Opportunities in Two-Dimensional Materials Beyond Graphene. *ACS Nano* **2013**, *7*, 2898-2926.

(10)   Das, S.; Robinson, J. A.; Dubey, M.; Terrones, H.; Terrones, M. Beyond Graphene: Progress in Novel Two-Dimensional Materials and van der Waals Solids. *Annual Review of Materials Research* **2015**, *45*, 1-27.

(11)   Li, L. K.; Yu, Y. J.; Ye, G. J.; Ge, Q. Q.; Ou, X. D.; Wu, H.; Feng, D. L.; Chen, X. H.; Zhang, Y. B. Black phosphorus field-effect transistors. *Nat Nanotechnol* **2014**, *9*, 372-377.

(12)   Ling, X.; Wang, H.; Huang, S.; Xia, F.; Dresselhaus, M. S. The renaissance of black phosphorus. *Proceedings of the National Academy of Sciences* **2015**, *112*, 4523-4530.

(13)   Liu, H.; Du, Y.; Deng, Y.; Ye, P. D. Semiconducting black phosphorus: synthesis, transport properties and electronic applications. *Chemical Society Reviews* **2015**, *44*, 2732-2743.

(14)   Pumera, M.; Sofer, Z. 2D Monoelemental Arsenene, Antimonene, and Bismuthene: Beyond Black Phosphorus. *Advanced Materials* **2017**, 1605299-n/a.

(15)   Tao, L.; Cinquanta, E.; Chiappe, D.; Grazianetti, C.; Fanciulli, M.; Dubey, M.; Molle, A.; Akinwande, D. Silicene field-effect transistors operating at room temperature. *Nat Nano* **2015**, *10*, 227-231.

(16)   Mannix, A. J.; Zhou, X.-F.; Kiraly, B.; Wood, J. D.; Alducin, D.; Myers, B. D.; Liu, X.; Fisher, B. L.; Santiago, U.; Guest, J. R.; Yacaman, M. J.; Ponce, A.; Oganov, A. R.; Hersam, M. C.; Guisinger, N. P. Synthesis of borophenes: Anisotropic, two-dimensional boron polymorphs. *Science* **2015**, *350*, 1513-1516.

(17)   Bandurin, D. A.; Tyurnina, A. V.; Yu, G. L.; Mishchenko, A.; Zólyomi, V.; Morozov, S. V.; Kumar, R. K.; Gorbachev, R. V.; Kudrynskyi, Z. R.; Pezzini, S.; Kovalyuk, Z. D.; Zeitler, U.; Novoselov, K. S.;

Patanè, A.; Eaves, L.; Grigorieva, I. V.; Fal'ko, V. I.; Geim, A. K.; Cao, Y. High electron mobility, quantum Hall effect and anomalous optical response in atomically thin InSe. *Nat Nano* **2017**, *12*, 223-227.

(18)     Tongay, S.; Sahin, H.; Ko, C.; Luce, A.; Fan, W.; Liu, K.; Zhou, J.; Huang, Y.-S.; Ho, C.-H.; Yan, J.; Ogletree, D. F.; Aloni, S.; Ji, J.; Li, S.; Li, J.; Peeters, F. M.; Wu, J. Monolayer behaviour in bulk ReS2 due to electronic and vibrational decoupling. *Nature Communications* **2014**, *5*, 3252.

(19)     Yoon, Y.; Ganapathi, K.; Salahuddin, S. How Good Can Monolayer MoS2 Transistors Be? *Nano Letters* **2011**, *11*, 3768-3773.

(20)     Fiori, G.; Bonaccorso, F.; Iannaccone, G.; Palacios, T.; Neumaier, D.; Seabaugh, A.; Banerjee, S. K.; Colombo, L. Electronics based on two-dimensional materials. *Nat Nano* **2014**, *9*, 768-779.

(21)     Alam, K.; Lake, R. K. Monolayer MoS$_2$ Transistors Beyond the Technology Road Map. *IEEE Transactions on Electron Devices* **2012**, *59*, 3250-3254.

(22)     Fivaz, R.; Mooser, E. Electron-Phonon Interaction in Semiconducting Layer Structures. *Physical Review* **1964**, *136*, A833-A836.

(23)     Fivaz, R.; Mooser, E. Mobility of Charge Carriers In Semiconducting Layer Structures. *Physical Review* **1967**, *163*, 743-755.

(24)     Wilson, J. A.; Yoffe, A. D. The transition metal dichalcogenides discussion and interpretation of the observed optical, electrical and structural properties. *Advances in Physics* **1969**, *18*, 193-335.

(25)     Kuc, A.; Zibouche, N.; Heine, T. Influence of Quantum Confinement on the Electronic Structure of the Transition Metal Sulfide TS2. *Physical Review B* **2011**, *83*, 245213.

(26)     Kaasbjerg, K.; Thygesen, K. S.; Jacobsen, K. W. Phonon-limited mobility in n-type single-layer MoS$_2$ from first principles. *Physical Review B* **2012**, *85*, 115317.

(27)     Wang, K.-C.; Stanev, T. K.; Valencia, D.; Charles, J.; Henning, A.; Sangwan, V. K.; Lahiri, A.; Mejia, D.; Sarangapani, P.; Povolotskyi, M. Control of interlayer delocalization in 2H transition metal dichalcogenides. *arXiv preprint arXiv:1703.02191* **2017**.

(28)     Dean, C. R.; Young, A. F.; MericI; LeeC; WangL; SorgenfreiS; WatanabeK; TaniguchiT; KimP; Shepard, K. L.; HoneJ. Boron nitride substrates for high-quality graphene electronics. *Nat Nano* **2010**, *5*, 722-726.

(29)     Cui, X.; Lee, G.-H.; Kim, Y. D.; Arefe, G.; Huang, P. Y.; Lee, C.-H.; Chenet, D. A.; Zhang, X.; Wang, L.; Ye, F.; Pizzocchero, F.; Jessen, B. S.; Watanabe, K.; Taniguchi, T.; Muller, D. A.; Low, T.; Kim, P.; Hone, J. Multi-terminal transport measurements of MoS2 using a van der Waals heterostructure device platform. *Nat Nano* **2015**, *10*, 534-540.

(30)     Kang, K.; Xie, S.; Huang, L.; Han, Y.; Huang, P. Y.; Mak, K. F.; Kim, C.-J.; Muller, D.; Park, J. High-mobility three-atom-thick semiconducting films with wafer-scale homogeneity. *Nature* **2015**, *520*, 656-660.

(31)     Britnell, L.; Gorbachev, R. V.; Jalil, R.; Belle, B. D.; Schedin, F.; Mishchenko, A.; Georgiou, T.; Katsnelson, M. I.; Eaves, L.; Morozov, S. V.; Peres, N. M. R.; Leist, J.; Geim, A. K.; Novoselov, K. S.; Ponomarenko, L. A. Field-effect tunneling transistor based on vertical graphene heterostructures. *Science* **2012**, *335*, 947-950.

(32)     Britnell, L.; Ribeiro, R. M.; Eckmann, A.; Jalil, R.; Belle, B. D.; Mishchenko, A.; Kim, Y.-J.; Gorbachev, R. V.; Georgiou, T.; Morozov, S. V.; Grigorenko, A. N.; Geim, A. K.; Casiraghi, C.; Neto, A. H. C.; Novoselov, K. S. Strong light-matter interactions in heterostructures of atomically thin films. *Science* **2013**, *340*, 1311-1314

(33)     Grigorieva, I. V.; Geim, A. K. Van der Waals heterostructures. *Nature* **2013**, *499*, 419-425.

(34)     Withers, F.; Del Pozo-Zamudio, O.; Mishchenko, A.; Rooney, A. P.; Gholinia, A.; Watanabe, K.; Taniguchi, T.; Haigh, S. J.; Geim, A. K.; Tartakovskii, A. I.; Novoselov, K. S. Light-emitting diodes by band-structure engineering in van der Waals heterostructures. *Nat Mater* **2015**, *14*, 301-306.

(35)     Buscema, M.; Island, J. O.; Groenendijk, D. J.; Blanter, S. I.; Steele, G. A.; van der Zant, H. S. J.; Castellanos-Gomez, A. Photocurrent generation with two-dimensional van der Waals semiconductors. *Chemical Society Reviews* **2015**, *44*, 3691-3718.

(36)     Lotsch, B. V. Vertical 2D Heterostructures. *Annual Review of Materials Research* **2015**, *45*, 85-109.

(37)     Yang, H.; Heo, J.; Park, S.; Song, H. J.; Seo, D. H.; Byun, K.-E.; Kim, P.; Yoo, I.; Chung, H.-J.; Kim, K. Graphene Barristor, A Triode Device with a Gate-Controlled Schottky Barrier. *Science* **2012**, *336*, 1140-1143.

(38)     Jariwala, D.; Marks, T. J.; Hersam, M. C. Mixed-dimensional van der Waals heterostructures. *Nature Materials* **2016**, DOI: 10.1038/NMAT4703.

(39)     Jariwala, D.; Sangwan, V. K.; Wu, C.-C.; Prabhumirashi, P. L.; Geier, M. L.; Marks, T. J.; Lauhon, L. J.; Hersam, M. C. Gate-Tunable Carbon Nanotube–MoS$_2$ Heterojunction p-n Diode. *Proceedings of the National Academy of Sciences* **2013**, *110*, 18076–18080.

(40)     Sze, S. M.; Ng, K. K.: *Physics of Semiconductor Devices*; Wiley-Interscience, 2006.

(41)     Shastry, T. A.; Balla, I.; Bergeron, H.; Amsterdam, S. H.; Marks, T. J.; Hersam, M. C. Mutual Photoluminescence Quenching and Photovoltaic Effect in Large-Area Single-Layer MoS2–Polymer Heterojunctions. *ACS Nano* **2016**, *10*, 10573-10579.

(42)     Miao, J.; Zhang, S.; Cai, L.; Scherr, M.; Wang, C. Ultrashort Channel Length Black Phosphorus Field-Effect Transistors. *ACS Nano* **2015**, *9*, 9236-9243.

(43)     Kim, I. S.; Sangwan, V. K.; Jariwala, D.; Wood, J. D.; Park, S.; Chen, K.-S.; Shi, F.; Ruiz-Zepeda, F.; Ponce, A.; Jose-Yacaman, M.; Dravid, V. P.; Marks, T. J.; Hersam, M. C.; Lauhon, L. J. Influence of Stoichiometry on the Optical and Electrical Properties of Chemical Vapor Deposition Derived MoS2. *ACS Nano* **2014**, *8*, 10551-10558.

(44)     Wood, J. D.; Wells, S. A.; Jariwala, D.; Chen, K.-S.; Cho, E.; Sangwan, V. K.; Liu, X.; Lauhon, L. J.; Marks, T. J.; Hersam, M. C. Effective Passivation of Exfoliated Black Phosphorus Transistors against Ambient Degradation. *Nano Letters* **2014**, *14*, 6964-6970.

(45)     Allain, A.; Kang, J.; Banerjee, K.; Kis, A. Electrical contacts to two-dimensional semiconductors. *Nat Mater* **2015**, *14*, 1195-1205.

(46)     Koppens, F. H. L.; Mueller, T.; Avouris, P.; Ferrari, A. C.; Vitiello, M. S.; Polini, M. Photodetectors based on graphene, other two-dimensional materials and hybrid systems. *Nat Nano* **2014**, *9*, 780-793.

(47)     Ji, Q.; Zhang, Y.; Zhang, Y.; Liu, Z. Chemical vapour deposition of group-VIB metal dichalcogenide monolayers: engineered substrates from amorphous to single crystalline. *Chemical Society Reviews* **2015**, *44*, 2587-2602.

(48)     Kang, J.; Sangwan, V. K.; Wood, J. D.; Hersam, M. C. Solution-Based Processing of Monodisperse Two-Dimensional Nanomaterials. *Accounts of Chemical Research* **2017**, *50*, 943-951.

(49)     Ryder, C. R.; Wood, J. D.; Wells, S. A.; Hersam, M. C. Chemically Tailoring Semiconducting Two-Dimensional Transition Metal Dichalcogenides and Black Phosphorus. *ACS Nano* **2016**, *10*, 3900-3917.

(50)     Dean, C.; Young, A. F.; Wang, L.; Meric, I.; Lee, G. H.; Watanabe, K.; Taniguchi, T.; Shepard, K.; Kim, P.; Hone, J. Graphene based heterostructures. *Solid State Communications* **2012**, *152*, 1275-1282.

(51)     Adler, D. Mechanisms for Metal-Nonmental Transitions in Transition-Metal Oxides and Sulfides. *Reviews of Modern Physics* **1968**, *40*, 714-736.

(52)     Wilson, J. A.; Di Salvo, F. J.; Mahajan, S. Charge-density waves and superlattices in the metallic layered transition metal dichalcogenides. *Advances in Physics* **1975**, *24*, 117-201.

(53)     Castro Neto, A. H.; Guinea, F.; Peres, N. M. R.; Novoselov, K. S.; Geim, A. K. The electronic properties of graphene. *Reviews of Modern Physics* **2009**, *81*, 109-162.

(54)     Das Sarma, S.; Adam, S.; Hwang, E. H.; Rossi, E. Electronic transport in two-dimensional graphene. *Reviews of Modern Physics* **2011**, *83*, 407-470.

(55)     Frindt, R. F. Single Crystals of MoS2 Several Molecular Layers Thick. *Journal of Applied Physics* **1966**, *37*, 1928-1929.

(56)     Joensen, P.; Frindt, R. F.; Morrison, S. R. Single-layer MoS2. *Materials Research Bulletin* **1986**, *21*, 457-461.

(57)     Kappera, R.; Voiry, D.; Yalcin, S. E.; Branch, B.; Gupta, G.; Mohite, A. D.; Chhowalla, M. Phase-engineered low-resistance contacts for ultrathin MoS2 transistors. *Nat Mater* **2014**, *13*, 1128-1134.

(58)     Kanazawa, T.; Amemiya, T.; Ishikawa, A.; Upadhyaya, V.; Tsuruta, K.; Tanaka, T.; Miyamoto, Y. Few-layer HfS2 transistors. *Scientific Reports* **2016**, *6*, 22277.

(59)     Yu, Y.; Yang, F.; Lu, X. F.; Yan, Y. J.; ChoYong, H.; Ma, L.; Niu, X.; Kim, S.; Son, Y.-W.; Feng, D.; Li, S.; Cheong, S.-W.; Chen, X. H.; Zhang, Y. Gate-tunable phase transitions in thin flakes of 1T-TaS2. *Nat Nano* **2015**, *10*, 270-276.

(60)     Bhimanapati, G. R.; Lin, Z.; Meunier, V.; Jung, Y.; Cha, J.; Das, S.; Xiao, D.; Son, Y.; Strano, M. S.; Cooper, V. R.; Liang, L.; Louie, S. G.; Ringe, E.; Zhou, W.; Kim, S. S.; Naik, R. R.; Sumpter, B. G.; Terrones, H.; Xia, F.; Wang, Y.; Zhu, J.; Akinwande, D.; Alem, N.; Schuller, J. A.; Schaak, R. E.; Terrones, M.; Robinson, J. A. Recent Advances in Two-Dimensional Materials beyond Graphene. *ACS Nano* **2015**, *9*, 11509-11539.

(61)     Mak, K. F.; He, K.; Lee, C.; Lee, G. H.; Hone, J.; Heinz, T. F.; Shan, J. Tightly Bound Trions in Monolayer $MoS_2$. *Nature materials* **2012**, *12*, 207-211.

(62)     Mak, K. F.; He, K.; Shan, J.; Heinz, T. F. Control of Valley Polarization in Monolayer $MoS_2$ by Optical Helicity. *Nature Nanotechnology* **2012**, *7*, 494-498.

(63)     Xia, F.; Wang, H.; Xiao, D.; Dubey, M.; Ramasubramaniam, A. Two-dimensional material nanophotonics. *Nat Photon* **2014**, *8*, 899-907.

(64)     Jung, C. S.; Shojaei, F.; Park, K.; Oh, J. Y.; Im, H. S.; Jang, D. M.; Park, J.; Kang, H. S. Red-to-Ultraviolet Emission Tuning of Two-Dimensional Gallium Sulfide/Selenide. *ACS Nano* **2015**, *9*, 9585-9593.

(65)     Ahn, J.-H.; Lee, M.-J.; Heo, H.; Sung, J. H.; Kim, K.; Hwang, H.; Jo, M.-H. Deterministic Two-Dimensional Polymorphism Growth of Hexagonal n-Type SnS2 and Orthorhombic p-Type SnS Crystals. *Nano Letters* **2015**, *15*, 3703-3708.

(66)     Hasan, M. Z.; Kane, C. L. Colloquium. *Reviews of Modern Physics* **2010**, *82*, 3045-3067.

(67)     Thomas, J.; Jezequel, G.; Pollini, I. Optical properties of layered transition-metal halides. *Journal of Physics: Condensed Matter* **1990**, *2*, 5439.

(68)     McGuire, M. A.; Dixit, H.; Cooper, V. R.; Sales, B. C. Coupling of Crystal Structure and Magnetism in the Layered, Ferromagnetic Insulator CrI3. *Chemistry of Materials* **2015**, *27*, 612-620.

(69)     Balendhran, S.; Walia, S.; Nili, H.; Ou, J. Z.; Zhuiykov, S.; Kaner, R. B.; Sriram, S.; Bhaskaran, M.; Kalantar-zadeh, K. Two-Dimensional Molybdenum Trioxide and Dichalcogenides. *Advanced Functional Materials* **2013**, *23*, 3952-3970.

(70)     Ma, R.; Sasaki, T. Two-Dimensional Oxide and Hydroxide Nanosheets: Controllable High-Quality Exfoliation, Molecular Assembly, and Exploration of Functionality. *Accounts of Chemical Research* **2015**, *48*, 136-143.

(71)     Naguib, M.; Mochalin, V. N.; Barsoum, M. W.; Gogotsi, Y. 25th Anniversary Article: MXenes: A New Family of Two-Dimensional Materials. *Advanced Materials* **2014**, *26*, 992-1005.

(72)     Cao, D. H.; Stoumpos, C. C.; Farha, O. K.; Hupp, J. T.; Kanatzidis, M. G. 2D Homologous Perovskites as Light-Absorbing Materials for Solar Cell Applications. *Journal of the American Chemical Society* **2015**, *137*, 7843-7850.

(73)     Mannix, A. J.; Kiraly, B.; Hersam, M. C.; Guisinger, N. P. Synthesis and chemistry of elemental 2D materials. *Nature Reviews Chemistry* **2017**, *1*, 0014.

(74)     Ridley, B. K.: *Quantum Processes in Semiconductors*; Oxford University Press, 2013.

(75)     Gong, C.; Zhang, H.; Wang, W.; Colombo, L.; Wallace, R. M.; Cho, K. Band alignment of two-dimensional transition metal dichalcogenides: Application in tunnel field effect transistors. *Applied Physics Letters* **2013**, *103*, 053513.

(76)     Kuc, A.; Heine, T. The electronic structure calculations of two-dimensional transition-metal dichalcogenides in the presence of external electric and magnetic fields. *Chemical Society Reviews* **2015**, *44*, 2603-2614.

(77)     Klein, J.; Wierzbowski, J.; Regler, A.; Becker, J.; Heimbach, F.; Müller, K.; Kaniber, M.; Finley, J. J. Stark Effect Spectroscopy of Mono- and Few-Layer MoS2. *Nano Letters* **2016**, *16*, 1554-1559.

(78)     Chu, T.; Ilatikhameneh, H.; Klimeck, G.; Rahman, R.; Chen, Z. Electrically Tunable Bandgaps in Bilayer MoS2. *Nano Letters* **2015**, *15*, 8000-8007.

(79)     Lu, C.-P.; Li, G.; Mao, J.; Wang, L.-M.; Andrei, E. Y. Bandgap, Mid-Gap States, and Gating Effects in MoS2. *Nano Letters* **2014**, *14*, 4628-4633.

(80)     Mak, K. F.; McGill, K. L.; Park, J.; McEuen, P. L. The valley Hall effect in MoS$_2$ transistors. *Science* **2014**, *344*, 1489-1492.

(81)     Far-Infrared Cyclotron Resonance Absorptions in Black Phosphorus Single Crystals. *Journal of the Physical Society of Japan* **1983**, *52*, 3544-3553.

(82)     Lin, Y.-C.; Komsa, H.-P.; Yeh, C.-H.; Björkman, T.; Liang, Z.-Y.; Ho, C.-H.; Huang, Y.-S.; Chiu, P.-W.; Krasheninnikov, A. V.; Suenaga, K. Single-Layer ReS2: Two-Dimensional Semiconductor with Tunable In-Plane Anisotropy. *ACS Nano* **2015**, *9*, 11249-11257.

(83)     Xia, F.; Wang, H.; Jia, Y. Rediscovering black phosphorus as an anisotropic layered material for optoelectronics and electronics. *Nature Communications* **2014**, *5*, 4458.

(84)     Keyes, R. W. The Electrical Properties of Black Phosphorus. *Physical Review* **1953**, *92*, 580-584.

(85)     Das, S.; Appenzeller, J. Where Does the Current Flow in Two-Dimensional Layered Systems? *Nano Letters* **2013**, *13*, 3396-3402.

(86)     Das, S.; Chen, H.-Y.; Penumatcha, A. V.; Appenzeller, J. High Performance Multilayer MoS2 Transistors with Scandium Contacts. *Nano Letters* **2012**, *13*, 100-105.

(87)     Ma, N.; Jena, D. Charge Scattering and Mobility in Atomically Thin Semiconductors. *Physical Review X* **2014**, *4*, 011043.

(88)     Trolle, M. L.; Pedersen, T. G.; Véniard, V. Model dielectric function for 2D semiconductors including substrate screening. *Scientific Reports* **2017**, *7*, 39844.

(89)     Scholz, A.; Stauber, T.; Schliemann, J. Plasmons and screening in a monolayer of MoS$_2$. *Physical Review B* **2013**, *88*, 035135.

(90)     Lin, Y.; Ling, X.; Yu, L.; Huang, S.; Hsu, A. L.; Lee, Y.-H.; Kong, J.; Dresselhaus, M. S.; Palacios, T. Dielectric Screening of Excitons and Trions in Single-Layer MoS2. *Nano Letters* **2014**, *14*, 5569-5576.

(91)     Nan, M.; Debdeep, J. Carrier statistics and quantum capacitance effects on mobility extraction in two-dimensional crystal semiconductor field-effect transistors. *2D Materials* **2015**, *2*, 015003.

(92)     Bao, W.; Cai, X.; Kim, D.; Sridhara, K.; Fuhrer, M. S. High mobility ambipolar MoS2 field-effect transistors: Substrate and dielectric effects. *Applied Physics Letters* **2013**, *102*, 042104.

(93)     Ghatak, S.; Pal, A. N.; Ghosh, A. Nature of Electronic States in Atomically Thin MoS2 Field-Effect Transistors. *ACS Nano* **2011**, *5*, 7707-7712.

(94)     Sun, Y.; Wang, R.; Liu, K. Substrate induced changes in atomically thin 2-dimensional semiconductors: Fundamentals, engineering, and applications. *Applied Physics Reviews* **2017**, *4*, 011301.

(95)     Kim, S.; Konar, A.; Hwang, W. S.; Lee, J. H.; Lee, J.; Yang, J.; Jung, C.; Kim, H.; Yoo, J. B.; Choi, J. Y.; Jin, Y. W.; Lee, S. Y.; Jena, D.; Choi, W.; Kim, K. High-Mobility and Low-Power Thin-Film Transistors Based on Multilayer MoS$_2$ Crystals. *Nature communications* **2012**, *3*, 1011.

(96)     Jariwala, D.; Sangwan, V. K.; Late, D. J.; Johns, J. E.; Dravid, V. P.; Marks, T. J.; Lauhon, L. J.; Hersam, M. C. Band-like transport in high mobility unencapsulated single-layer MoS$_2$ transistors. *Applied Physics Letters* **2013**, *102*, 173107-173104.

(97)     Baugher, B. W. H.; Churchill, H. O. H.; Yang, Y.; Jarillo-Herrero, P. Intrinsic Electronic Transport Properties of High-Quality Monolayer and Bilayer MoS2. *Nano Letters* **2013**, *13*, 4212-4216.

(98)     Radisavljevic, B.; Kis, A. Mobility engineering and a metal–insulator transition in monolayer MoS2. *Nat Mater* **2013**, *12*, 815-820.

(99)     Sangwan, V. K.; Arnold, H. N.; Jariwala, D.; Marks, T. J.; Lauhon, L. J.; Hersam, M. C. Low-Frequency Electronic Noise in Single-Layer MoS2 Transistors. *Nano Letters* **2013**, *13*, 4351-4355.

(100)    Ong, Z.-Y.; Fischetti, M. V. Mobility enhancement and temperature dependence in top-gated single-layer MoS$_2$. *Physical Review B* **2013**, *88*, 165316.

(101)    Lauer, I.; Antoniadis, D. A. Enhancement of electron mobility in ultrathin-body silicon-on-insulator MOSFETs with uniaxial strain. *IEEE Electron Device Letters* **2005**, *26*, 314-316.

(102)    Yee Chia, Y.; Subramanian, V.; Kedzierski, J.; Peiqi, X.; Tsu-Jae, K.; Bokor, J.; Chenming, H. Nanoscale ultra-thin-body silicon-on-insulator P-MOSFET with a SiGe/Si heterostructure channel. *IEEE Electron Device Letters* **2000**, *21*, 161-163.

(103)    Poljak, M.; Jovanovic, V.; Grgec, D.; Suligoj, T. Assessment of Electron Mobility in Ultrathin-Body InGaAs-on-Insulator MOSFETs Using Physics-Based Modeling. *IEEE Transactions on Electron Devices* **2012**, *59*, 1636-1643.

(104)    Yuan, Y.; Giri, G.; Ayzner, A. L.; Zoombelt, A. P.; Mannsfeld, S. C. B.; Chen, J.; Nordlund, D.; Toney, M. F.; Huang, J.; Bao, Z. Ultra-high mobility transparent organic thin film transistors grown by an off-centre spin-coating method. *Nature Communications* **2014**, *5*, 3005.

(105)    Barquinha, P.; Pereira, L.; Gonçalves, G.; Martins, R.; Fortunato, E. Toward High-Performance Amorphous GIZO TFTs. *Journal of The Electrochemical Society* **2009**, *156*, H161-H168.

(106)    Sangwan, V. K.; Ortiz, R. P.; Alaboson, J. M. P.; Emery, J. D.; Bedzyk, M. J.; Lauhon, L. J.; Marks, T. J.; Hersam, M. C. Fundamental Performance Limits of Carbon Nanotube Thin-Film Transistors Achieved Using Hybrid Molecular Dielectrics. *ACS Nano* **2012**, *6*, 7480-7488.

(107)    Desai, S. B.; Madhvapathy, S. R.; Sachid, A. B.; Llinas, J. P.; Wang, Q.; Ahn, G. H.; Pitner, G.; Kim, M. J.; Bokor, J.; Hu, C.; Wong, H.-S. P.; Javey, A. MoS$_2$ transistors with 1-nanometer gate lengths. *Science* **2016**, *354*, 99-102.

(108)    Fiori, G.; Szafranek, B. N.; Iannaccone, G.; Neumaier, D. Velocity saturation in few-layer MoS2 transistor. *Applied Physics Letters* **2013**, *103*, 233509.

(109)    Zhang, F.; Appenzeller, J. Tunability of Short-Channel Effects in MoS2 Field-Effect Devices. *Nano Letters* **2015**, *15*, 301-306.

(110)    Liu, H.; Neal, A. T.; Ye, P. D. Channel Length Scaling of MoS2 MOSFETs. *ACS Nano* **2012**, *6*, 8563-8569.

(111)    Sanne, A.; Ghosh, R.; Rai, A.; Yogeesh, M. N.; Shin, S. H.; Sharma, A.; Jarvis, K.; Mathew, L.; Rao, R.; Akinwande, D.; Banerjee, S. Radio Frequency Transistors and Circuits Based on CVD MoS2. *Nano Letters* **2015**, *15*, 5039-5045.

(112)    Wang, H.; Wang, X.; Xia, F.; Wang, L.; Jiang, H.; Xia, Q.; Chin, M. L.; Dubey, M.; Han, S.-j. Black Phosphorus Radio-Frequency Transistors. *Nano Letters* **2014**, *14*, 6424-6429.

(113)    Gong, C.; Colombo, L.; Wallace, R. M.; Cho, K. The Unusual Mechanism of Partial Fermi Level Pinning at Metal–MoS2 Interfaces. *Nano Letters* **2014**, *14*, 1714-1720.

(114)    Chen, J.-R.; Odenthal, P. M.; Swartz, A. G.; Floyd, G. C.; Wen, H.; Luo, K. Y.; Kawakami, R. K. Control of Schottky Barriers in Single Layer MoS2 Transistors with Ferromagnetic Contacts. *Nano Letters* **2013**, *13*, 3106-3110.

(115)    Wang, L.; Meric, I.; Huang, P. Y.; Gao, Q.; Gao, Y.; Tran, H.; Taniguchi, T.; Watanabe, K.; Campos, L. M.; Muller, D. A.; Guo, J.; Kim, P.; Hone, J.; Shepard, K. L.; Dean, C. R. One-Dimensional Electrical Contact to a Two-Dimensional Material. *Science* **2013**, *342*, 614-617.

(116)    Chuang, S.; Battaglia, C.; Azcatl, A.; McDonnell, S.; Kang, J. S.; Yin, X.; Tosun, M.; Kapadia, R.; Fang, H.; Wallace, R. M.; Javey, A. MoS2 P-type Transistors and Diodes Enabled by High Work Function MoOx Contacts. *Nano Letters* **2014**, *14*, 1337-1342.

(117)    Mouri, S.; Miyauchi, Y.; Matsuda, K. Tunable Photoluminescence of Monolayer MoS2 via Chemical Doping. *Nano Letters* **2013**, *13*, 5944-5948.

(118)    van der Zande, A. M.; Huang, P. Y.; Chenet, D. A.; Berkelbach, T. C.; You, Y.; Lee, G.-H.; Heinz, T. F.; Reichman, D. R.; Muller, D. A.; Hone, J. C. Grains and grain boundaries in highly crystalline monolayer molybdenum disulphide. *Nat Mater* **2013**, *12*, 554-561.

(119)    Bettis Homan, S.; Sangwan, V. K.; Balla, I.; Bergeron, H.; Weiss, E. A.; Hersam, M. C. Ultrafast Exciton Dissociation and Long-Lived Charge Separation in a Photovoltaic Pentacene–MoS2 van der Waals Heterojunction. *Nano Letters* **2017**, *17*, 164-169.

(120)    Nan, H.; Wang, Z.; Wang, W.; Liang, Z.; Lu, Y.; Chen, Q.; He, D.; Tan, P.; Miao, F.; Wang, X.; Wang, J.; Ni, Z. Strong Photoluminescence Enhancement of MoS2 through Defect Engineering and Oxygen Bonding. *ACS Nano* **2014**, *8*, 5738-5745.

(121)    Amani, M.; Lien, D.-H.; Kiriya, D.; Xiao, J.; Azcatl, A.; Noh, J.; Madhvapathy, S. R.; Addou, R.; KC, S.; Dubey, M.; Cho, K.; Wallace, R. M.; Lee, S.-C.; He, J.-H.; Ager, J. W.; Zhang, X.; Yablonovitch, E.; Javey, A. Near-unity photoluminescence quantum yield in MoS$_2$. *Science* **2015**, *350*, 1065-1068.

(122)    Azizi, A.; Zou, X.; Ercius, P.; Zhang, Z.; Elías, A. L.; Perea-López, N.; Stone, G.; Terrones, M.; Yakobson, B. I.; Alem, N. Dislocation motion and grain boundary migration in two-dimensional tungsten disulphide. *Nature Communications* **2014**, *5*, 4867.

(123)    Yu, Z. G.; Zhang, Y.-W.; Yakobson, B. I. An Anomalous Formation Pathway for Dislocation-Sulfur Vacancy Complexes in Polycrystalline Monolayer MoS2. *Nano Letters* **2015**, *15*, 6855-6861.

(124)    Sangwan, V. K.; Jariwala, D.; Kim, I. S.; Chen, K.-S.; Marks, T. J.; Lauhon, L. J.; Hersam, M. C. Gate-tunable memristive phenomena mediated by grain boundaries in single-layer MoS2. *Nature Nanotechnology* **2015**, *10*, 403-406.

(125)    Komsa, H.-P.; Krasheninnikov, A. V. Two-Dimensional Transition Metal Dichalcogenide Alloys: Stability and Electronic Properties. *The Journal of Physical Chemistry Letters* **2012**, *3*, 3652-3656.

(126)    Gong, Y.; Liu, Z.; Lupini, A. R.; Shi, G.; Lin, J.; Najmaei, S.; Lin, Z.; Elías, A. L.; Berkdemir, A.; You, G.; Terrones, H.; Terrones, M.; Vajtai, R.; Pantelides, S. T.; Pennycook, S. J.; Lou, J.; Zhou, W.; Ajayan, P. M. Band Gap Engineering and Layer-by-Layer Mapping of Selenium-Doped Molybdenum Disulfide. *Nano Letters* **2014**, *14*, 442-449.

(127)    Chen, Y.; Xi, J.; Dumcenco, D. O.; Liu, Z.; Suenaga, K.; Wang, D.; Shuai, Z.; Huang, Y.-S.; Xie, L. Tunable Band Gap Photoluminescence from Atomically Thin Transition-Metal Dichalcogenide Alloys. *ACS Nano* **2013**, *7*, 4610-4616.

(128)    Lu, A.-Y.; Zhu, H.; Xiao, J.; Chuu, C.-P.; Han, Y.; Chiu, M.-H.; Cheng, C.-C.; Yang, C.-W.; Wei, K.-H.; Yang, Y.; Wang, Y.; Sokaras, D.; Nordlund, D.; Yang, P.; Muller, D. A.; Chou, M.-Y.; Zhang, X.; Li, L.-J. Janus monolayers of transition metal dichalcogenides. *Nat Nano* **2017**, *advance online publication*.

(129)    Yan, R.; Fathipour, S.; Han, Y.; Song, B.; Xiao, S.; Li, M.; Ma, N.; Protasenko, V.; Muller, D. A.; Jena, D.; Xing, H. G. Esaki Diodes in van der Waals Heterojunctions with Broken-Gap Energy Band Alignment. *Nano Letters* **2015**, *15*, 5791-5798.

(130)    Li, M.-Y.; Shi, Y.; Cheng, C.-C.; Lu, L.-S.; Lin, Y.-C.; Tang, H.-L.; Tsai, M.-L.; Chu, C.-W.; Wei, K.-H.; He, J.-H.; Chang, W.-H.; Suenaga, K.; Li, L.-J. Epitaxial growth of a monolayer WSe$_2$-MoS$_2$ lateral p-n junction with an atomically sharp interface. *Science* **2015**, *349*, 524-528.

(131)    Roy, T.; Tosun, M.; Cao, X.; Fang, H.; Lien, D.-H.; Zhao, P.; Chen, Y.-Z.; Chueh, Y.-L.; Guo, J.; Javey, A. Dual-Gated MoS2/WSe2 van der Waals Tunnel Diodes and Transistors. *ACS Nano* **2015**, *9*, 2071-2079.

(132)    Sarkar, D.; Xie, X.; Liu, W.; Cao, W.; Kang, J.; Gong, Y.; Kraemer, S.; Ajayan, P. M.; Banerjee, K. A subthermionic tunnel field-effect transistor with an atomically thin channel. *Nature* **2015**, *526*, 91-95.

(133)    Lee, C.-H.; Lee, G.-H.; van der Zande, A. M.; Chen, W.; Li, Y.; Han, M.; Cui, X.; Arefe, G.; Nuckolls, C.; Heinz, T. F.; Guo, J.; Hone, J.; Kim, P. Atomically thin p–n junctions with van der Waals heterointerfaces. *Nat Nano* **2014**, *9*, 676-681.

(134)    Nourbakhsh, A.; Zubair, A.; Dresselhaus, M. S.; Palacios, T. Transport Properties of a MoS2/WSe2 Heterojunction Transistor and Its Potential for Application. *Nano Letters* **2016**, *16*, 1359-1366.

(135)    Jariwala, D.; Howell, S. L.; Chen, K.-S.; Kang, J.; Sangwan, V. K.; Filippone, S. A.; Turrisi, R.; Marks, T. J.; Lauhon, L. J.; Hersam, M. C. Hybrid, gate-tunable, van der Waals p-n heterojunctions from pentacene and MoS$_2$. *Nano Letters* **2016**, *16*, 497–503.

(136)    Zhou, R.; Ostwal, V.; Appenzeller, J. Vertical versus Lateral Two-Dimensional Heterostructures: On the Topic of Atomically Abrupt p/n-Junctions. *Nano Letters* **2017**, *17*, 4787–4792.

(137)    Jariwala, D.; Sangwan, V. K.; Seo, J.-W. T.; Xu, W.; Smith, J.; Kim, C. H.; Lauhon, L. J.; Marks, T. J.; Hersam, M. C. Large-Area, Low-Voltage, Antiambipolar Heterojunctions from Solution-Processed Semiconductors. *Nano Letters* **2015**, *15*, 416-421.

(138)    Prins, F.; Goodman, A. J.; Tisdale, W. A. Reduced Dielectric Screening and Enhanced Energy Transfer in Single- and Few-Layer MoS2. *Nano Letters* **2014**, *14*, 6087-6091.

(139)    Tan, C.; Zhang, H. Two-dimensional transition metal dichalcogenide nanosheet-based composites. *Chemical Society Reviews* **2015**, *44*, 2713-2731.

(140)    Zhu, J.; Hersam, M. C. Assembly and Electronic Applications of Colloidal Nanomaterials. *Advanced Materials* **2017**, *29*, 1603895.

(141)    Kelly, A. G.; Hallam, T.; Backes, C.; Harvey, A.; Esmaeily, A. S.; Godwin, I.; Coelho, J.; Nicolosi, V.; Lauth, J.; Kulkarni, A.; Kinge, S.; Siebbeles, L. D. A.; Duesberg, G. S.; Coleman, J. N. All-printed thin-film transistors from networks of liquid-exfoliated nanosheets. *Science* **2017**, *356*, 69-73.

(142)    Kang, J.; Sangwan, V. K.; Wood, J. D.; Liu, X.; Balla, I.; Lam, D.; Hersam, M. C. Layer-by-Layer Sorting of Rhenium Disulfide via High-Density Isopycnic Density Gradient Ultracentrifugation. *Nano Letters* **2016**, *16*, 7216-7223.

(143)    Kang, J.; Wood, J. D.; Wells, S. A.; Lee, J.-H.; Liu, X.; Chen, K.-S.; Hersam, M. C. Solvent Exfoliation of Electronic-Grade, Two-Dimensional Black Phosphorus. *ACS Nano* **2015**, *9*, 3596-3604.

(144)    Favron, A.; Gaufres, E.; Fossard, F.; Phaneuf-Lheureux, A.-L.; Tang, N. Y. W.; Levesque, P. L.; Loiseau, A.; Leonelli, R.; Francoeur, S.; Martel, R. Photooxidation and quantum confinement effects in exfoliated black phosphorus. *Nat Mater* **2015**, *14*, 826-832.

(145)    Ryder, C. R.; Wood, J. D.; Wells, S. A.; Yang, Y.; Jariwala, D.; Marks, T. J.; Schatz, G. C.; Hersam, M. C. Covalent functionalization and passivation of exfoliated black phosphorus via aryl diazonium chemistry. *Nat Chem* **2016**, *8*, 597-602.

(146)    Zhao, M.; Ye, Y.; Han, Y.; Xia, Y.; Zhu, H.; Wang, S.; Wang, Y.; Muller, D. A.; Zhang, X. Large-scale chemical assembly of atomically thin transistors and circuits. *Nat Nano* **2016**, *11*, 954-959.

(147)    Wu, J.; Yuan, H.; Meng, M.; Chen, C.; Sun, Y.; Chen, Z.; Dang, W.; Tan, C.; Liu, Y.; Yin, J.; Zhou, Y.; Huang, S.; Xu, H. Q.; Cui, Y.; Hwang, H. Y.; Liu, Z.; Chen, Y.; Yan, B.; Peng, H. High electron mobility and quantum oscillations in non-encapsulated ultrathin semiconducting Bi2O2Se. *Nat Nano* **2017**, *12*, 530-534.

(148)    Ray, K.; Yore, A. E.; Mou, T.; Jha, S.; Smithe, K. K. H.; Wang, B.; Pop, E.; Newaz, A. K. M. Photoresponse of Natural van der Waals Heterostructures. *ACS Nano* **2017**, *11*, 6024-6030.

(149)    Peng, B.; Ang, P. K.; Loh, K. P. Two-dimensional dichalcogenides for light-harvesting applications. *Nano Today* **2015**, *10*, 128-137.

(150)    Kang, J.; Tongay, S.; Zhou, J.; Li, J.; Wu, J. Band offsets and heterostructures of two-dimensional semiconductors. *Applied Physics Letters* **2013**, *102*, 012111.

(151)    Huang, Z.; Zhang, W.; Zhang, W. Computational Search for Two-Dimensional MX2 Semiconductors with Possible High Electron Mobility at Room Temperature. *Materials* **2016**, *9*, 716.